\documentclass[
 reprint,
superscriptaddress,
 amsmath,amssymb,
 aps,
prb,
]{revtex4-2}

\usepackage{graphicx}
\usepackage[center]{}

\usepackage{fancyhdr}

\usepackage{psfrag}
\usepackage{bm}

\fancypagestyle{FirstPage}{
\lfoot{Copyright (2022) R. Finn, S. Schulz. This article is distributed under a Creative Commons Attribution (CC BY) License.} 
}

\begin{document}

\preprint{APS/123-QED}

\title{Impact of random alloy fluctuations on the electronic and optical properties of (Al,Ga)N quantum wells: Insights from tight-binding calculations}

\author{Robert Finn}
 \email{robert.finn@tyndall.ie}
 \affiliation{Tyndall National Institute, University College Cork,
Cork, T12 R5CP, Ireland}
\author{Stefan Schulz}
\affiliation{Department of Physics, University College Cork, Cork, T12 YN60, Ireland}
\affiliation{Tyndall National Institute, University College Cork,
Cork, T12 R5CP, Ireland}

\date{\today}

\begin{abstract}
Light emitters based on the semiconductor alloy aluminium gallium nitride ((Al,Ga)N) have gained significant attention in recent years due to their potential for a wide range of applications in the ultraviolet (UV) spectral window. However, current state-of-the-art (Al,Ga)N light emitters exhibit very low internal quantum efficiencies (IQEs). Therefore, understanding the fundamental electronic and optical properties of (Al,Ga)N-based quantum wells is key to improving the IQE. Here, we target the electronic and optical properties of c-plane Al$_x$Ga$_{1-x}$N/AlN quantum wells by means of an empirical atomistic tight-binding model. Special attention is paid to the impact of random alloy fluctuations on the results as well as the aluminium content x in the well. We find that across the studied Al content range (from 10\% to 75\% Al) strong hole wave function localization effects are observed. Additionally, with increasing Al content, electron wave functions start also to exhibit carrier localization features. Overall, our investigations on the electronic structure of c-plane Al$_x$Ga$_{1-x}$N/AlN quantum wells reveal that already random alloy fluctuations are sufficient to lead to (strong) carrier localization effects. Furthermore, our results indicate that random alloy fluctuations impact the degree of optical polarization in c-plane Al$_x$Ga$_{1-x}$N quantum wells. We find that the switching from transverse electric to transverse magnetic light polarization occurs at higher Al contents in the atomistic calculation, which accounts for random alloy fluctuations, when compared to the outcome of widely used virtual crystal approximations. This observation is important for light extraction efficiencies in (Al,Ga)N-based light emitting diodes operating in the deep UV.  
\end{abstract}

\maketitle

\section{Introduction}

\thispagestyle{FirstPage}

Alloy disorder induced carrier localization effects in semiconductor materials have been a topic of extensive research for several decades~\cite{BrKe2003,Thou1970}. In the field of nitride-based semiconductor heterostructures it has gained significant momentum in recent years, due to the importance of such effects for the electronic and optical properties of 'classical' (e.g. light emitting diodes (LEDs)) and non-classical light sources (e.g. entangled photon emitters)~\cite{PiLi2017,DiVPe2020,ODoLu2021,PaSc2020,PaSc2021}. It has been shown, both in experiment and theory, that the semiconductor alloy indium gallium nitride ((In,Ga)N) is particularly prone to alloy induced carrier localization effects. Such carrier localization effects are key to explaining for instance the defect insensitivity and thus the high efficiency of (In,Ga)N-based LEDs operating in the blue to violet spectral range~\cite{ChUe2006}. Here, it is widely accepted that alloy fluctuations localize carriers and thus keep them away from defects, which can act as non-radiative recombination centres~\cite{ChUe2006}. Recently, the III-N alloy aluminium gallium nitride ((Al,Ga)N) has attracted considerable attention due to its potential for a wide range of applications in the ultraviolet (UV) spectral region~\cite{AmCo2020_JPD,KnSe2019}. Thanks to their ultra-wide-band gaps, (Al,Ga)N alloys and quantum wells (QWs) with high Al contents are particularly attractive for deep UV applications such as water purification or sterilisation procedures~\cite{AmCo2020_JPD,KnSe2019}.

Experimental studies on (Al,Ga)N QWs give clear indications of strong carrier localization effects in these systems, e.g. shown by an ''S-shaped'' temperature dependence of the photoluminescence (PL) peak position~\cite{FrNi2020_JAP} or the very large full width at half maximum (FWHM) PL linewidth~\cite{FrNi2020_PSS}. 

Theoretical studies focusing on the impact of alloy fluctuations on carrier localization effects in (Al,Ga)N QW systems are sparse. Recent calculations using modified continuum-based (single-band effective mass) descriptions have targeted pure GaN wells with (Al,Ga)N barriers, thus neglecting any alloy fluctuations \emph{inside} the well~\cite{RoPa2019}. In the continuum-based studies of Rudinsky and Karpov~\cite{RuKa2020} it is \emph{assumed} that hole states are strongly localized but electron states are delocalized. It is important to note that in Ref.~\onlinecite{RuKa2020} the carrier localization characteristics are an \emph{input} to the model.
There exists currently no atomistic theoretical calculation that investigates the impact of alloy fluctuations on the electronic and optical properties of (Al,Ga)N/AlN \emph{QWs}.

We close this gap here by employing an empirical atomistic tight-binding (TB) model that accounts for alloy fluctuations on a microscopic level. This approach has already been benchmarked against experimental results for \emph{bulk} (Al,Ga)N systems showing, for instance, good agreement in the band gap energy as a function of the alloy content~\cite{CoSc2015}. Equipped with this model, our calculations on (Al,Ga)N/AlN QWs reveal that alloy fluctuations lead to strong hole wave function localization effects, independent of the here studied Al contents, namely 10\%, 25\%, 50\% and 75\%. Our theoretical studies also give indications of \emph{electron} localization effects due to alloy fluctuations at higher Al-contents (e.g. $\geq$ 50\%). In (In,Ga)N systems, electron wave-functions are mainly localized by well width fluctuations, and to a lesser extent by alloy fluctuations~\cite{TaMc2018}. We attribute this fundamental difference between (Al,Ga)N and (In,Ga)N alloys to differences in, for instance, the conduction band effective masses of InN, GaN and AlN, as we will discuss in more detail below. These electron localization effects, along with localization effects introduced by well width fluctuations (see Ref.~\cite{RoPa2019}) may now also explain the experimental observation that carrier localization effects in (Al,Ga)N/AlN QWs can be even stronger when compared to (In,Ga)N systems~\cite{FrNi2020_JAP}. Furthermore, our theoretical studies indicate that when considering alloy fluctuations in the calculations, the cross-over from transverse electric (TE) to transverse magnetic (TM) polarization occurs at higher Al contents when compared to the outcome of a virtual crystal approximation (VCA) that neglects alloy fluctuations. Thus, while alloy fluctuations may be detrimental for the electron and hole wave function overlap, they can be beneficial for the light extraction efficiency (LEE) in deep UV light emitters, since the LEE in such emitters suffers from TM rather than TE polarized emission~\cite{AmCo2020_JPD,NaLi2004,KoKn2010,RyCh2013}.  

The paper is organized as follows: we describe in Sec.~\ref{sec:Theory} the theoretical framework along with information on the QW model systems. The results of our study are presented in Sec.~\ref{sec:results}. A summary and conclusion is given in Sec.~\ref{sec:Concl}.

\section{Theoretical Framework}
\label{sec:Theory}

In this section we describe the microscopic theoretical framework used to study the electronic and optical properties of (Al,Ga)N/AlN QWs. In Sec.~\ref{sec:TBmodel} we discuss briefly the TB model, the valence force field (VFF) method and the local polarization theory employed here. Subsequently, in Sec.~\ref{sec:QWModel}, the QW model systems as well as some general results on how alloy fluctuations affect strain and polarization fields in (Al,Ga)N/AlN QWs are presented.

\subsection{Tight-binding model, valence force field model and local polarization theory}
\label{sec:TBmodel}

In order to capture the impact of alloy fluctuations on the electronic and optical properties of III-N-based heterostructures, an electronic structure model with an atomistic resolution is ideally suited. However, the supercell size, i.e. the number of atoms involved, required to describe (Al,Ga)N/AlN QWs of realistic size and to capture carrier localization effects on an atomistic level is beyond the reach of standard density functional theory (DFT). To be able to (i) treat alloy fluctuations on an atomistic level and (ii) a large number of atoms, we employ here an empirical nearest neighbour $sp^3$ TB model. The employed model (including strain and polarization fields) has already been used to investigate the electronic structure of bulk (Al,Ga)N systems~\cite{CoSc2015} and builds on the model described in Ref.~\onlinecite{ScCa2015} for (In,Ga)N systems. In the following we will only briefly discuss the main ingredients of the simulation framework, and e.g. summarize TB model, VFF model and local polarization theory. More details can be found in Refs.~\onlinecite{CaSc2013local,ScCa2015,CoSc2015}. 
\subsubsection{TB model}
The TB parameters for the binary materials AlN and GaN are determined by fitting the respective $sp^3$ TB band structures to data obtained from hybrid-functional DFT calculations~\cite{ScCa2013_Apex,CaSc2013local,CoSc2015}. To obtain the TB parameters for (Al,Ga)N alloys, we proceed as follows. For the cations, (Al, Ga), the nearest neighbour environment always consists of nitrogen (N) atoms in the alloy. In this case, the TB parameters (onsite and hopping matrix elements) from the binaries are used. For N atoms, one is left with a varying number of Al and Ga atoms as nearest neighbors. For the on-site TB parameters, we use a weighted average (depending on the number of Al and Ga nearest neighbor atoms) of the binary AlN and GaN materials; for the hopping matrix elements values from the binary material are used. Overall, the employed procedure is a widely used approximation in the literature~\cite{OReLi2002,LiPo1992,BoKh2007}.

Given the differences in the lattice constants of wurtzite GaN and AlN (approximately 2.4\% for the $a$ lattice constant), (local) strain effects have to be taken into account.
These effects are included in the TB Hamiltonian via the Pikus-Bir Hamiltonian~\cite{WiSc2006,ScBa2010} as on-site corrections to the $s$- and $p$-orbital energies. The deformation potentials required in the Pikus-Bir Hamiltonian are taken again from hybrid functional DFT calculations~\cite{YaRi2009}. 

Finally, wurtzite GaN and AlN exhibit spontaneous and piezoelectric polarization vector fields~\cite{BeFi1997}; here both macroscopic and (local) microscopic built-in polarization fields have to be taken into account to obtain an accurate description of the electronic and optical properties of (Al,Ga)N/AlN QWs. To do so, we include the (local) built-in potential as a site-diagonal correction in the TB Hamiltonian, inline with Refs.~\onlinecite{SaAr2002,ScBa2012}.

Having discussed how to include strain and polarization fields in the TB Hamiltonian, we describe in the following how to calculate these contributions in systems with a  large number of atoms while still keeping an atomistic resolutions. We start with the strain field before turning to the built-in polarization fields.

\subsubsection{Valence force field model}

In order to describe (local) strain effects on an atomistic level in (Al,Ga)N alloys, but ultimately $c$-plane (Al,Ga)N/AlN QWs,  we apply the VFF model described in Refs.~\onlinecite{CaSc2013local,ScCa2015}. This model accounts for bond bending, bond stretching, and cross terms but also Coulomb effects. The latter is for instance required to obtain a good description of the $c/a$ ratio in wurtzite III-N systems. Overall, the required VFF parameters are obtained in the case of the binary materials by fitting to the lattice constants and the elastic tensor components from DFT calculations~\cite{CaSc2013local,CaSc2012}. Similarly to the TB parameters, averages of the VFF parameters are used in the alloy case. Strain field results for (Al,Ga)N/AlN QWs obtained within this framework are discussed in more detail in Sec.~\ref{sec:QWModel}.
Given the large number of atoms involved in describing the strain field in a QW system, a numerically efficient routine is required to minimize the VFF elastic energy. To achieve this, the VFF model has been implemented using the software package \textsc{lammps}~\cite{Plim1995}.

\subsubsection{Local polarization theory}

As already mentioned above, due to the symmetry of the underlying crystal structure, wurtzite AlN and GaN possess both spontaneous and strain dependent piezoelectric polarization vector fields~\cite{BeFi1997}. In the following we distinguish between microscopic and macroscopic effects. We classify macroscopic effects as mainly those contributions that are also found in standard continuum-based models (e.g. $\mathbf{k}\cdot\mathbf{p}$ methods) where the electrostatic built-in field mainly arises from a discontinuity in the polarization vector field at the interfaces of a heterostructure, e.g. at the interface between a $c$-plane (Al,Ga)N QW and an AlN barrier. However, when dealing with an (Al,Ga)N alloy in a QW, one is also left with local strain field fluctuations due to local alloy fluctuations. These local 'deformations' of the lattice result, for instance, in local piezoelectric polarization field fluctuations. To take both macroscopic and microscopic polarization field contributions into account, we employ the local polarization theory introduced in Ref.~\cite{CaSc2013local}; the required material parameters for the AlN and GaN are also given in Ref.~\cite{CaSc2013local}.
We will discuss the polarization fields in $c$-plane (Al,Ga)N/GaN QWs, obtained from our local polarization theory, in more detail below.

\subsection{Model quantum well system}
\label{sec:QWModel}

\begin{figure}
\includegraphics[width=1.05\columnwidth]{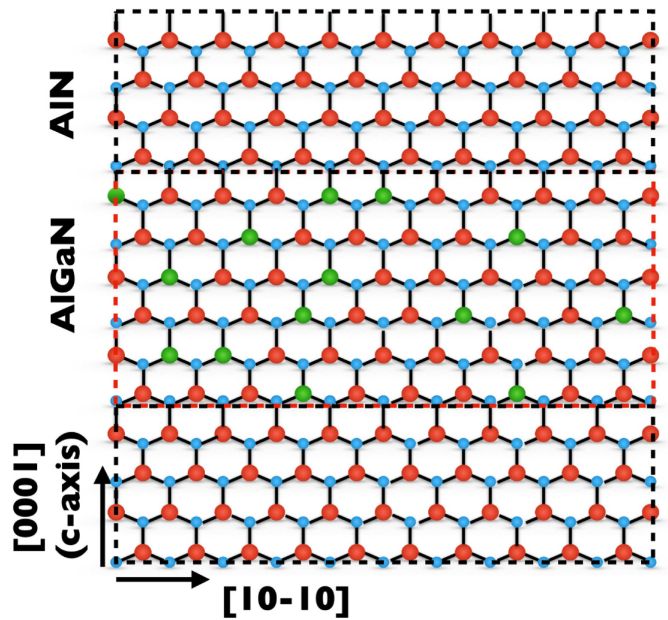}
\caption{Schematic illustration of the supercell underlying the atomistic tight-binding calculations of the electronic and optical properties of $c$-plane Al$_{x}$Ga$_{1-x}$N/AlN quantum wells. In the region indicated by 'AlGaN', we assume a random distribution of Al (green) and Ga (red) atoms.}
\label{fig:schermaticQW}
\end{figure}

Equipped with the information about the theoretical model, we describe in the following the QW model structures it is applied to. Given the microscopic resolution of our framework, we use supercells (with periodic boundary conditions) that contain 81,920 atoms, corresponding to a cell with the dimension of approximately 10 nm $\times$ 9 nm $\times$ 10 nm, in our calculations of the electronic and optical properties of $c$-plane Al$_x$Ga$_{1-x}$N/AlN QW systems.  In general, we start from a cell of pure AlN (so all atoms in the cell are either Al or N). In a second step, Al atoms are replaced by Ga atoms in  a specified sub-region of the cell which defines the well (between specified $z$-coordinates of the cell). The number of Al atoms replacing Ga atoms is determined by the Al/Ga content in the well. Here, no preferential positioning or clustering is assumed, building on findings from (In,Ga)N systems~\cite{RiBo2018}. A schematic illustration of the simulation cell, including random alloy fluctuations, is given in Fig.~\ref{fig:schermaticQW}. 

Once the (different) atoms are placed on this wurtzite lattice, again reflecting the targeted alloy content, the elastic (VFF) energy of the system is minimised. To do so, internal degrees of freedom for each atom are `optimised' while the simulation cell is only allowed to expand or contract along the $z$-axis ($c$-axis), thus assuming here pseudomorphic growth on AlN. After the equilibrium structure is obtained, the underlying atom coordinates can be used for the calculation of the (local) polarization fields and ultimately the TB model. 

Below we investigate systems with Al contents of 10\%, 25\%, 50\% and 75\%. The chosen alloy contents allow us to cover the range experimentally relevant for developing emitters spanning from the UV-A all the way into the deep UV-C spectral region. We note that in the present study we always select the same number of atomic planes in which Al atoms are replaced by Ga atoms. As a consequence, after relaxing the atomic positions, and especially for different Al contents, the well width between the different calculations may be slightly different. However, overall the QW width of the different systems studied is approximately 2.9 nm.

\begin{figure}
\begin{tabular}{c}
\includegraphics[width=\columnwidth]{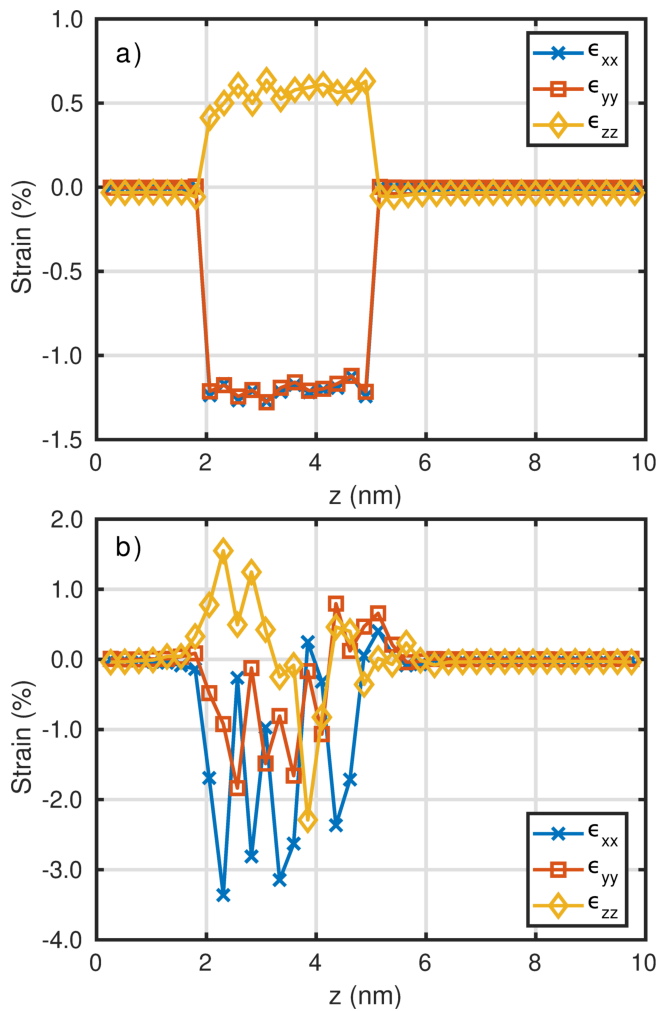}
\end{tabular}
\caption{Strain tensor components $\epsilon_{xx}$, $\epsilon_{yy}$ and $\epsilon_{zz}$ for a linescan along the $z$-axis ($c$-axis) of the simulation cell for an Al$_{0.5}$Ga$_{0.5}$N/AlN quantum well. Data averaged over the simulation cell  is shown in (a) while (b) shows an individual linescan.}
\label{fig:strain_QW}
\end{figure}

Using the approach discussed in Ref.~\cite{PrKi1998,ShTa2012}, Fig.~\ref{fig:strain_QW} displays the diagonal components of the strain tensor, $\epsilon_{xx}$, $\epsilon_{yy}$ and $\epsilon_{zz}$, for a linescan along the $c$-axis of a $c$-plane Al$_{0.5}$Ga$_{0.5}$N/AlN QW. The data in Fig.~\ref{fig:strain_QW} (a) are averaged over the supercell and reveal a behavior expected from a continuum-based description of a $c$-plane (Al,Ga)N/AlN QW~\cite{CaSc2011}: (i) there is no strain in the AlN barriers, (ii) due to the symmetry of the $c$-plane and the in-plane lattice mismtach between AlN and Al$_{0.5}$Ga$_{0.5}$N, to a first approximation $\epsilon_{xx}\approx\epsilon_{yy}\approx$ -1.2\% and (iii) $\epsilon_{zz}>0$. While Fig.~\ref{fig:strain_QW} (a) shows in general a similar strain profile as can be expected from a continuum-based model, the figure reveals also differences, namely that even the averaged data show some fluctuations. This stems from the fact that the strain fluctuates locally, due to local fluctuations in Al and Ga atoms. Figure~\ref{fig:schermaticQW} (b) displays this clearly. Here the strain tensor components $\epsilon_{ii}$ are shown for a \emph{single} linescan along the $c$-axis. In this case, $\epsilon_{xx}$ differs from $\epsilon_{yy}$ as well as $\epsilon_{zz}$; all these components may even change sign. Such a feature is not captured in "standard" continuum-based models or a virtual crystal approximation (VCA) of an alloy.

\begin{figure}
    \includegraphics[width=90mm]{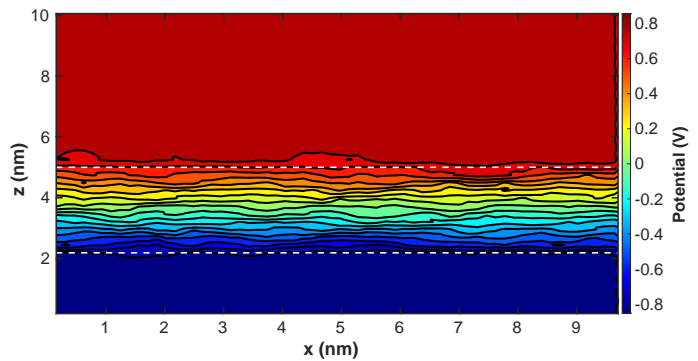}
\caption{Contour plot of the polarization potential $\phi_p$ arising from spontaneous and piezoelectric polarization for an arbitrarily chosen $x$-$z$-plane in a $c$-plane Al$_{0.5}$Ga$_{0.5}$N/AlN quantum well.}
\label{fig:PolPot}
\end{figure}

As already mentioned above, the overall strain profile, but also the (local) fluctuations, will give rise to (local) polarization fields. This situation is visible in Fig.~\ref{fig:PolPot} which depicts a contourplot of the electrostatic built-in potential $\phi_p$ arising from spontaneous and piezoelectric polarization in a $c$-plane Al$_{0.5}$Ga$_{0.5}$N/AlN QW. The local fluctuations in the built-in potential are reflected for instance in the situation that the countour lines within the QW are not parallel; parallel contourlines are usually expected from continuum-based calculations of polarization fields in III-N QW structures, justifying the 1-D calculations widely employed for these systems.

The analysis above already reveals that local alloy fluctuations noticeably affect strain fields and built-in potentials. Thus, these effects may significantly impact the electronic and ultimately the optical properties of (Al,Ga)N QWs. To gain insight into these questions, 20 different random alloy configurations have been generated for each Al content (10\%, 25\%, 50\% and 75\%) in the well, and the influence of the alloy microstructure on the results will be discussed in more detail below.

\section{Results}
\label{sec:results}

In the following section, Sec.~\ref{sec:GenCon}, we start with some general considerations of wurtzite III-N materials and focus on differences between InN, GaN and AlN and how those differences may affect alloy-induced carrier localization effects. In Sec~\ref{sec:SPS} we discuss the electronic structure before turning to optical properties in Sec.~\ref{sec:DOP} and specifically to the degree of optical polarization.  

\subsection{General considerations}
\label{sec:GenCon}

Before presenting our results on the electronic and optical properties of $c$-plane (Al,Ga)N QWs, we discuss fundamental properties of AlN, GaN and InN systems first. This information will allow us to understand similarities and differences between (Al,Ga)N and (In,Ga)N-based heterostructures. 
In the experimental studies by Frankerl~\emph{et al.}~\cite{FrNi2020_APL} it is discussed that the atomic radius of an Al atom, in contrast to an In atom, is very similar to a Ga atom. So at first glance, replacing an Al atom with a Ga atom should perturb the local ``energy landscape'' to a lesser extent in an (Al,Ga)N alloy than when replacing Ga by In atoms as in an (In,Ga)N alloy. However, and in addition to this observation, several further factors are important when examining the impact of alloy disorder on the electronic structure. Firstly, even though the lattice mismatch in an (Al,Ga)N alloy is smaller in comparison to (In,Ga)N, the local strain effects seen in Fig.~\ref{fig:strain_QW} (b) can still lead to noticeable fluctuations in the polarization potential, and thus the local quantum confinement. In turn this may lead to carrier localization effects. 

Furthermore, the energy gap difference between AlN and GaN ($\Delta E^\text{AlN,GaN}_g=E^\text{AlN}_g-E^\text{GaN}_g\approx 2.5$ eV) is comparable to that of GaN and InN ($\Delta E^\text{InN,GaN}_g=E^\text{GaN}_g-E^\text{InN}_g\approx 2.8$)~\cite{RiWi2008}; thus locally high Ga or Al contents in an (Al,Ga)N alloy may lead to significant changes in the local band gap values. Moreover, differences in the electronegtivities for Al and Ga atoms (on Pauling scale $\chi^\text{Al}_p=1.61$; $\chi^\text{Ga}_p=1.81$) are larger when comparing this to Ga and In atoms (on Pauling scale $\chi^\text{In}_p=1.78$)~\cite{Paul1960}. In alloys such as GaAsN, it has been shown that a contrast in electronegativity can lead to strong carrier localization effects and band gap changes~\cite{OReLi2004}. Thus taking all this together, and even though the atomic radii may not be too different between Al and Ga atoms, the above factors are indicative of strong carrier localization effects in (Al,Ga)N. 

Such an 'expectation' is further supported by the fact that in general the effective masses of the holes and the electrons in both AlN and GaN are close to, or even higher, when compared to InN~\cite{RiWi2008,ScBa2010}. Similar to an (In,Ga)N QW system, strong localization effects for holes can be expected in (Al,Ga)N wells, given the high and similar effective hole masses in both systems. However, the situation for electrons in (Al,Ga)N alloys may be different to (In,Ga)N since the conduction band effective mass in GaN and AlN is a factor of order 3 larger when compared to InN ($m^{\text{AlN},\parallel}_e=0.322$, $m^{\text{AlN},\perp}_e=0.329$; $m^{\text{GaN},\parallel}_e=0.186$, $m^{\text{GaN},\perp}_e=0.209$ and $m^{\text{InN},\parallel}_e=0.065$, $m^{\text{InN},\perp}_e=0.068$)~\cite{RiWi2008,PeFe2001,SuUe1995,KiLa1997}. Based on this one may expect a stronger impact of alloy and well width fluctuations on the electron wave function in (Al,Ga)N QWs in comparison to (In,Ga)N systems. This may especially be the case for systems with 50\% Al and 50\% Ga contents, where smaller local regions of 'pure' GaN may be surrounded by 'pure' AlN, which can generate a local GaN/AlN quantum dot. 

All this will contribute to the experimental observations that carrier localization effects seem to be more pronounced in (Al,Ga)N/AlN wells when compared to (In,Ga)N QW systems, especially for structures close to the 50\% Al and 50\% Ga case, as discussed in Refs.~\cite{FrNi2020_PSS,FrNi2020_APL,FrNi2020_JAP}. However, it is important to stress that a one-to-one comparison between (Al,Ga)N and (In,Ga)N  QWs is difficult, since high quality (In,Ga)N wells with 50\% In content can basically not being realised in experiment.      

\subsection{Single-particle energies and states}
\label{sec:SPS}

\begin{figure}[t!]
    \includegraphics[width=\columnwidth]{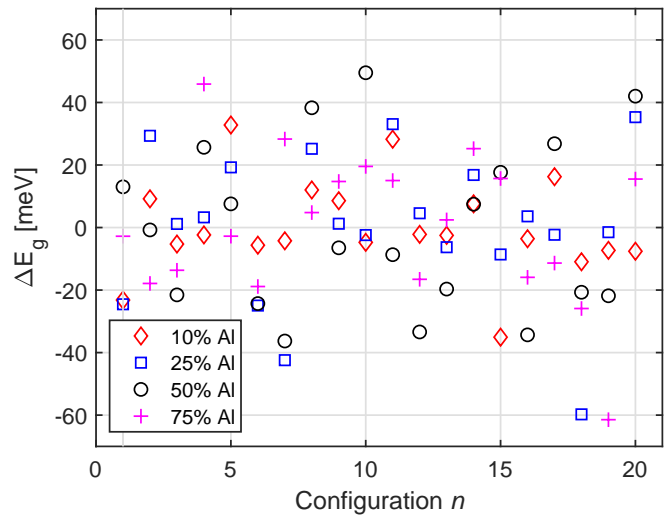}
\caption{Relative ground state transition energy \mbox{$\Delta E_g(n)=E_g(n)-E^\text{avg}_g$} in $c$-plane Al$_x$Ga$_{1-x}$N/AlN QWs for different microscopic alloy configurations $n$; more details are given in the main text. The data are shown for Al contents of 10\%, 25\%, 50\% and 75\% in the well.}
\label{fig:RelGSTran}
\end{figure}

We start our atomistic TB analysis of the electronic and optical properties of $c$-plane Al$_x$Ga$_{1-x}$N/AlN QWs, with Al contents ranging from 10\% up to 75\%, by looking at the \emph{average} ground state energy, \mbox{$E^\text{avg}_g=\frac{1}{N}\sum^N_{n}E_g(n)=\frac{1}{N}\sum^N_{n}E^e_\text{GS}(n)-E^h_\text{GS}(n)$} of the $N=20$ considered random alloy configurations per Al content. Here, $E^e_\text{GS}(n)$ and $E^h_\text{GS}(n)$ are the electron and hole ground state energy of a given alloy configuration $n$, respectively; note that the hole ground state energy corresponds to the valence ''band'' edge energy in the conduction valence band picture. For $E^\text{avg}_g$ we find on an absolute scale values of 2.40 $\pm$0.015 eV, 2.89 $\pm$0.024 eV, 3.85 $\pm$0.026 eV and 4.97 $\pm$0.023 eV for 10\%, 25\%, 50\% and 75\% Al in the well, respectively. Thus, our calculations reveal the expected behavior that with increasing Al composition in the well, $E^\text{avg}_g$ increases. 
Our values predicted for (Al,Ga)N/AlN QW systems on the high and low end of the studied Al contents are in good agreement with literature experimental data~\cite{AdSa2003, taLi2012, IcIw2014}. However, transition energies  reported in Refs.~\cite{FrNi2020_PSS, Fr2021} for wells with approximately 50\% Al content, appear to have  larger PL peak energies than the transition energy values calculated here. Differences may stem from well width fluctuations or screening of the internal built-in fields, to name only two potential sources. Further investigations into these aspects are required but are beyond the scope of this study, which focuses on the impact of random alloy fluctuations on the electronic and optical properties of $c$-plane (Al,Ga)N QWs. 

We note that the experimental investigations in Refs.~\onlinecite{FrNi2020_PSS,IcIw2014, taLi2012} all showed broad PL spectra. A similar situation has been observed in (In,Ga)N/GaN QW systems, where the broadening of the PL spectrum is usually attributed to alloy fluctuations and connected carrier localization effects~\cite{DaSc2016}.
To study the impact of alloy fluctuations on the electronic and optical properties of $c$-plane Al$_x$Ga$_{1-x}$N/AlN QWs within our atomistic model, we look at the \emph{relative} ground state transition energy \mbox{$\Delta E_g(n)=E_g(n)-E^\text{avg}_g$}. Thus we are analyzing by how much an individual configuration deviates from the average. The results of this analysis are displayed in Fig.~\ref{fig:RelGSTran} for the different alloy configurations and Al contents studied. One can infer from this figure that independent of the Al content, noticeable fluctuations in $\Delta E_g(n)$ even at low Al contents of 10\% are observed. Thus this finding is consistent with the experimental observation of broad PL spectra in (Al,Ga)N/AlN QW systems~\cite{FrNi2020_PSS,TaCa2016_RA,IcIw2014}. Moreover, and in line with our general considerations above, at 50\% and 75\% we find several extremely large deviations from the average transition energy and thus large $\Delta E_g(n)$ values (reaching values of $\approx \pm$ 60 meV).

\begin{figure}[t!]
\includegraphics[width=\columnwidth]{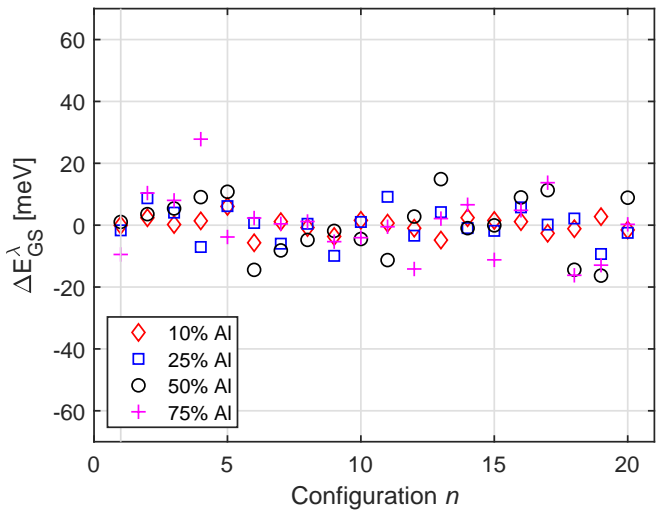}
\caption{Relative \emph{electron} ground state energy \mbox{$\Delta E^e_\text{GS}(n)=E^e_\text{GS}(n)-E^{e,\text{avg}}_\text{GS}$} in $c$-plane Al$_x$Ga$_{1-x}$N/AlN QWs for different microscopic configurations $n$; more details are given in the main text. The data are shown for Al contents of 10\%, 25\%, 50\% and 75\% in the well.}
\label{fig:RelEGS}
\end{figure}

This leaves the question of whether these fluctuations stem mainly from variations in hole, 
$E^h_\text{GS}$, or electron, $E^e_\text{GS}$, ground state energies. To shed light onto this question, we proceed with a similar approach as for the transition energies and calculate the relative variation $\Delta E^\lambda_\text{GS}=E^\lambda_\text{GS}(n)-E^{\lambda,\text{avg}}_\text{GS}$ of the ground state energies. Here $\lambda$ denotes electrons ($\lambda=e$) and holes ($\lambda=h$), respectively.

We start here with the electrons. Figure~\ref{fig:RelEGS} shows $\Delta E^\lambda_\text{GS}(n)$ for the different random alloy configurations $n$. Again the data are displayed for Al contents of 10\%, 25\%, 50\% and 75\% in the well. For the Al contents below 50\%, $\Delta E^e_\text{GS}(n)$ scatters between a maximum of $\pm 10$ meV. However, at higher Al content ($\geq 50$\%) noticeably larger $\Delta E^e_\text{GS}(n)$ values are observed. 
Based on our discussion in Sec~\ref{sec:GenCon}, one could expect that the electron wave functions are more strongly affected by alloy fluctuations at higher Al contents, due to local confinement effects.      

\begin{figure}[t!]
\includegraphics[width=\columnwidth]{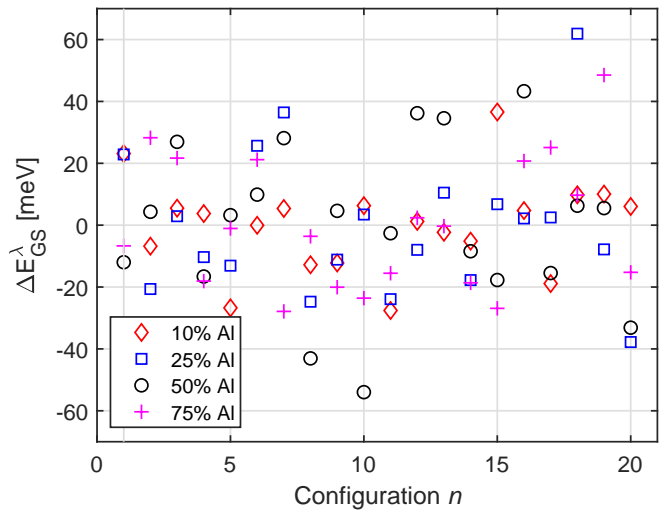}
\caption{Relative \emph{hole} ground state energy \mbox{$\Delta E^h_\text{GS}(n)=E^h_\text{GS}(n)-E^{h,\text{avg}}_\text{GS}$} in $c$-plane Al$_x$Ga$_{1-x}$N/AlN QWs for different microscopic configurations $n$; more details are given in the main text. The data are shown for Al contents of 10\%, 25\%, 50\% and 75\% in the well.}
\label{fig:RelHGS}
\end{figure}

The data for holes are shown in Fig.~\ref{fig:RelHGS}. In comparison to the electrons, see Fig.~\ref{fig:RelEGS}, the holes exhibit much larger fluctuations in $\Delta E^h_\text{GS}(n)$ as a function of the alloy configuration number $n$. Thus, one can conclude that especially for lower Al contents, the spread in the transition energies arises to a large extent from variations in the hole ground state energies.

Overall, the finding that the alloy microstructure has a stronger impact on the hole ground state energy is similar to (In,Ga)N systems~\cite{TaCa2016_RA}. We attributed this here to the fact that holes have a larger effective mass then electrons (even though the electron effective mass is increased in (Al,Ga)N when compared to (In,Ga)N) and can thus be localized in smaller regions with higher Ga contents. Having discussed ground state \emph{energies}, in a next step we focus on wave function localization characteristics.

\begin{figure*}
    \includegraphics[width=1.0\textwidth]{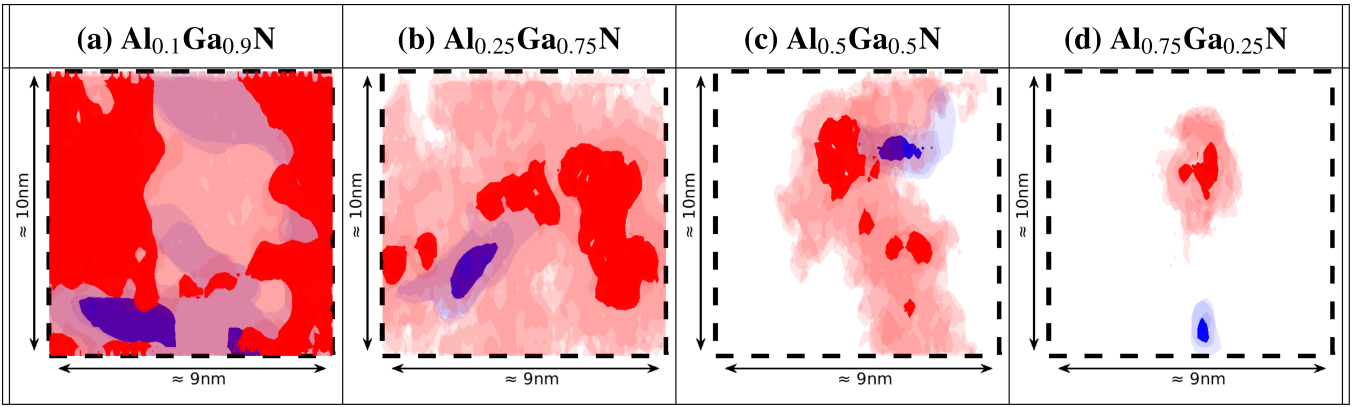}
\caption{Isosurface plots of the electron (red) and hole (blue) ground state charge densities for $c$-plane (Al,Ga)N/AlN QWs with (a) 10\%, (b) 25\%, (c) 50\% and (d) 75\% Al content. The light and dark surfaces correspond to 10\% and 50\% of the maximum values, respectively. The charge densities are shown for a top-view (down the $c$-axis). The dashed lines are used to indicate the supercell boundaries.}
\label{fig:ChargeDens}
\end{figure*}

Figure~\ref{fig:ChargeDens} displays isosurface plots of the electron (red) and hole (blue) ground state charge densities of selected alloy configurations for (a) 10\%,  (b) 25\%, (c) 50\% and (d) 75\% Al contents. For 10\% Al we have chosen configuration (Config.) 5 which gives a noticeable deviation from the average ground state transition energy (see Fig.~\ref{fig:RelGSTran}); this will give insight in how, even at lower Al contents, wave function localization effects can play a role. For 25\% Al we have chosen  Config. 3 and for 50\% Al Config. 2. These configurations are close to the respective average transition energies (see Fig.~\ref{fig:RelGSTran}) and will shed light on wave function localization effects in an 'average' configuration. Finally, for 75\% Al we have chosen Config. 18 in which both hole and electron ground state energies deviate noticeably from the average values (see Figs.~\ref{fig:RelEGS} and~\ref{fig:RelHGS}).  We note that Fig.~\ref{fig:ChargeDens} displays top-views of the charge density, thus looking down the $c$-axis ($z$-axis) of our simulation cell; the wave functions are spatially separated along the growth direction due to the presence of the electrostatic built-in field, which is not visible in the top-view. The light and dark isosurfaces correspond to 10\% and 50\% of the respective maximum charge density values. 

Looking at the QW system with 10\% Al first, Fig.~\ref{fig:ChargeDens} (a), one observes hole localization effects, whereas the electron wave function is more delocalized, thus more evenly distributed over the QW plane. We note that well width fluctuations or alloy fluctuations in the barrier material may introduce additional localization centres for electron wave functions~\cite{RoPa2019}. In terms of the carrier localization features, the here found situation is similar to (In,Ga)N/GaN systems~\cite{TaMc2018}. However, one needs to be careful with this comparisons. The studied Al$_{0.1}$Ga$_{0.9}$N/AlN well contains 90\% of the low band gap material. A high equality (In,Ga)N/GaN QW with such as high composition of the low band gap material (InN) cannot be grown. Moving on to the QW with 25\% Al, from Fig.~\ref{fig:ChargeDens} (b) one can infer that the hole is still strongly localized. However, the electron wave function is more noticeably affected by the alloy fluctuations in comparison to the 10\% Al case (see Fig.~\ref{fig:ChargeDens} (a)). This effect is even more pronounced in the 50\% Al case, see Fig.~\ref{fig:ChargeDens} (c). We note that for the 50\% Al case, we have chosen a configuration close to the average transition energy. Thus the observed electron and hole localization features are not an 'outlier' or an extreme configuration. We find in 11 out of the 20 configurations considered a very similar behavior to the situation depicted in Fig.~\ref{fig:ChargeDens} (c). The remaining 9 configurations exhibit a more delocalized electron charge density similar to the QW with 25\% Al as shown in Fig.~\ref{fig:ChargeDens} (b). Turning in the last step towards the QW system with 75\% Al, Fig.~\ref{fig:ChargeDens} (d) clearly shows that random alloy fluctuations can be sufficient to introduce localization effects for \emph{both} hole and electron charge densities. The feature of strong electron wave function localization, introduced purely by alloy fluctuations and not well width fluctuations, is usually not found in (In,Ga)N/GaN QWs~\cite{TaMc2018}. However, we highlight that for the chosen configuration both the electron and hole ground state energies deviate noticeably from the respective average values. Thus, the localization effect for the electrons is not seen in all configurations. In fact we observe this strong impact of the alloy fluctuations on the electron wave function in approximately 8 out of the 20 different alloy configurations. With the other configurations looking more akin to the 50\% sample in Fig.~\ref{fig:ChargeDens} (c).

As mentioned already above, in addition to these alloy induced carrier localization effects, the intrinsic electrostatic built-in field in nitride-based QWs leads to the situation that electrons and holes are spatially separated along the growth direction. Thus, the carriers localize on opposite well interfaces: the electrons at the upper (AlN barrier on (Al,Ga)N well) while the holes on the lower interface ((Al,Ga)N well on AlN barrier). This effect is often discussed in connection to the the quantum confined Stark effect (QCSE)~\cite{Re2016}. Overall, this aspect is now also important for alloy induced carrier localization effects: given that electrons and holes localize on opposite well interfaces, the alloy microstructure is different at these interfaces. Thus, and confirmed by Figs.~\ref{fig:ChargeDens}, electron and hole wave functions do not even localize at the same \emph{in-plane} position in the well, and may even be independently localized as observed in (In,Ga)N wells~\cite{DaSc2016}. Therefore, and in addition to the out-of plane spatial separation of electron and hole wave functions due to the QCSE, which is accounted for by standard continuum-based calculations, our calculations reveal that the carriers are also spatially separated in the growth plane. Overall, all these effects can have a detrimental impact on the radiative recombination rates.

In summary, our results show that random alloy fluctuations are already sufficient to facilitate strong carrier localization effects. However, so far we only studied ground state properties of (Al,Ga)N/AlN QWs. For optoelectronic devices such as LEDs, at elevated temperatures or carrier densities, not only the ground states but also the energetically higher/lower lying conduction/valence states are important. Localization characteristics in such excited states play an important role for quantities such as recombination rates (both radiative~\cite{NiKa2016,McMTa2020} and non-radiative~\cite{McMKi2022,JoTe2017}) but also the degree of optical polarization (DOP)~\cite{ScCa2013_Apex}. The latter aspect is of particular importance and interest for the LEE in deep UV light emitters~\cite{AmCo2020_JPD,KnKo2010,RyCh2013}. Therefore, we turn our attention in the following towards localization effects in excited states (energetically higher/lower lying states). 

\begin{figure}[t!]
    \includegraphics[width=\columnwidth]{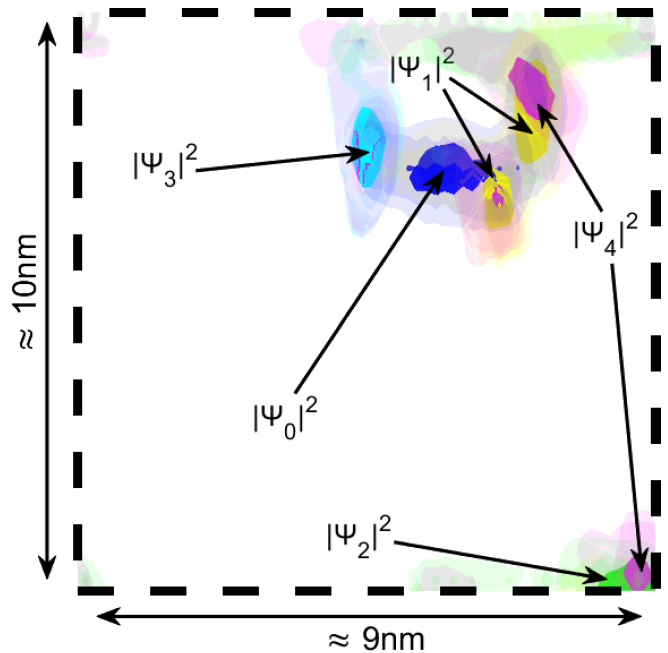}
    \caption{Isosurface plots of the charge densities of the five energetically lowest hole states ($\psi^h_0$ to $\psi^h_4$) in a $c$-plane (Al,Ga)N/AlN quantum well with 50\% Al content. The same alloy configuration as in Fig.~\ref{fig:ChargeDens} (c) is chosen here. The light and dark isosurfaces correspond to 10\% and 50\% of the respective maximum charge density values.}
\label{fig:CDExcitedHole}
\end{figure}

For electrons we find that while ground states may exhibit localization effects, excited states are more delocalized in nature (not shown). However, for holes the situation is different. Figure~\ref{fig:CDExcitedHole} displays isosurface plots of charge densities of the 5 energetically lowest hole states for the alloy configuration already studied in Fig.~\ref{fig:ChargeDens} (c) of a $c$-plane (Al,Ga)N QW with 50\% Al. 

When looking at Fig.~\ref{fig:CDExcitedHole}, it is clearly visible that not only the hole ground state is strongly localized but also the excited states. Also, we find that the charge densities may localize in spatially different regions.

Thus alloy fluctuations do not only affect ground but also the excited states. Moreover these alloy fluctuation induced localization features are also expected to impact the band or orbital character mixing in these excited states. As a consequence the DOP, $\rho$, can also be affected by these effects. We study this question, given its importance for the LEE in (Al,Ga)N-based deep UV light emitters, in more detail in the following section.

\subsection{Degree of optical polarization}
\label{sec:DOP}
    
As already discussed above, (Al,Ga)N based deep UV-emitters suffer from low LEE~\cite{AmCo2020_JPD}. A significant factor contributing to this issue stems from the polarization of the emitted light, which is tightly linked to fundamental differences in the electronic band structures of AlN and GaN~\cite{SaRe2011,ScCa2013_Apex,ScBa2010}. Of considerable interest here is the magnitude and sign of the crystal field splitting energy, $\Delta_\text{cf}$, which in the absence of spin-orbit coupling can be defined as the energy difference between the two topmost non-degenerate valence bands at $\mathbf{k}=\mathbf{0}$. For GaN, values of $+9$ to $+38$ meV have been reported in the literature, while for AlN much larger and \emph{negative} numbers, ranging from $-169$ meV up to $-295$ meV, are found~\cite{ScBa2010,VuMe2003,YaRi2011}. This difference in the sign of $\Delta_\text{cf}$ leads to a different ordering of the valence bands in GaN when compared to AlN.

When neglecting the weak spin-orbit coupling, the topmost valence band in AlN is mainly $p_z$-like in character~\cite{SaRe2011,ScCa2014}. This means that the band is a superposition of atomic-like $p_z$ basis states, which are oriented along the wurtzite $c$-axis. As a consequence, emission processes involving the ($s$-like) conduction band and this $p_z$-like valence band result in transverse magnetic (TM) polarised light emission. Due to the opposite sign of $\Delta_\text{cf}$ in GaN when compared to AlN, the topmost valence band in GaN is a linear combination of $p_x$- and $p_y$-like states. So in the case of GaN, and when only the top most valence band is involved in the light emission process, transverse electric (TE) polarized emission is expected and observed. This means in (Al,Ga)N alloys, and consequently (Al,Ga)N QWs, the optical polarization characteristics of the emitted light will switch from, for instance, TE to TM polarized light at a certain Al content. Since most of the conventional (Al,Ga)N LEDs are designed to be bottom or top emitting devices, TE polarization light emission is essential for efficient light extraction~\cite{NoCh2012,Ry2014}. However, for deep UV LEDs, high Al contents (e.g. $>$ 40 \%) are required to achieve the emission wavelength in the desired UV window~\cite{PeKi2010}. Consequently, to have both the desired wavelength and efficient light extraction, the cross-over from TE to TM should be shifted to Al contents as high as possible.

To analyze the optical polarization switching in more detail, we study similar to Ref.~\cite{NaLi2004}, the orbital character of the involved hole/valence states in our $c$-plane (Al,Ga)N QWs as a function of the Al content, $x$, but also temperature, $T$. To do so we define the temperature dependent DOP, $\rho(T)$, as follows:
\begin{equation}
 \rho(T)=\frac{\sum_if(E_i,T)(I^i_x+I^i_y)-\sum_if(E_i,T)I^i_z}{\sum_if(E_i,T)(I^i_x+I^i_y)+\sum_i f(E_i,T)I^i_z}\,\, .
 \label{eq:DOLP}
\end{equation}
Here $I^i_\alpha$ is the fraction of $p_x$ ($\alpha=x$), $p_y$ ($\alpha=y$) or $p_z$ ($\alpha=z$) orbital contribution in the $i$th QW state with the energy $E_i$; $f(E_i,T)$ denotes the Fermi function. Using Eq.~(\ref{eq:DOLP}), a value of $\rho=1$ corresponds to TE polarized light, while $\rho=-1$ indicates TM polarization.

Before looking at the DOP, we start with analyzing the orbital contributions in the hole ground and first four excited states for a QW with 50\% and 75\% Al, respectively; the data are displayed in Table~\ref{tab:bandmix} for the alloy configurations for which the ground state charge densities are depicted in Fig.~\ref{fig:ChargeDens} (c) and (d), respectively. Table~\ref{tab:bandmix} reveals that for an Al content of 50\%, these 5 states are dominated by $p_x$ and $p_y$-like orbital contributions. Thus, at least for lower temperatures and carrier densities, predominantly TE polarized light ($\rho>0$) can be expected, since in this case only states near the band edge (lowest lying hole states) are being populated. We note that due to the alloy disorder, the symmetry between $p_x$ and $p_y$-like orbitals is broken. Table~\ref{tab:bandmix} shows also that the $p_z$-orbital contribution in the first 5 hole states in a $c$-plane (Al,Ga)N/AlN QW with 75\% Al is strongly increased and may even be the largest contribution, e.g. states $\psi_1$, $\psi_2$ and $\psi_3$. However, even in that case $p_x$ and $p_y$ still remain significant. Thus the situation for the DOP is less clear cut in this situation and unpolarized or only weakly polarized light may be expected ($\rho\approx 0$).  

To gain insight into the impact of alloy fluctuations on the DOP, we use a VCA description of the $c$-plane (Al,Ga)N QW systems studied above as reference. This means instead of resolving the individual Ga and Al atoms in the well, all atoms in the QW are replaced by virtual (Al,Ga)N atoms. The required TB parameters for each Al content $x$ studied are determined by a linear interpolation of the TB parameters for GaN and AlN. Thus local fluctuations in the Al and Ga content are not accounted for and the VCA model is similar to a standard continuum-based description of an (Al,Ga)N QW.

\begin{table}[t!]
\caption{Orbital contributions (in \%) of the energetically lowest first five hole states $\psi_0-\psi_4$ of a $c$-plane (Al,Ga)N/AlN well with 50\% Al (upper part of table) and 75\% Al. We have chosen as an example the alloy configurations for which the hole ground state charge densities are depicted in Fig.~\ref{fig:ChargeDens} (c) and (d), respectively. }
\label{tab:bandmix}

\begin{tabular}{| c | c | c | c | c| c|}
\hline
 & \multicolumn{5}{c|}{\textbf{Al$_{0.5}$Ga$_{0.5}$N}}\\ \hline
     & $\psi^h_0$ & $\psi^h_1$ & $\psi^h_2$ & $\psi^h_3$ & $\psi^h_4$ \\ \hline
 $p_x$ (\%) &  22.43 &  62.84 &  22.17 &  68.64 &  59.39 \\
 $p_y$ (\%) &  76.48 &  35.62 &  76.49 &  28.32 &  38.36  \\
 $p_z$ (\%) &  0.95 &  1.38 &  1.19 &  2.79 &  2.06  \\
 $s$ (\%)   &  0.14 &  0.16 &  0.16 &  0.25 &  0.18 \\
\hline\hline
 & \multicolumn{5}{c|}{\textbf{Al$_{0.75}$Ga$_{0.25}$N}}\\ \hline
     & $\psi^h_0$ & $\psi^h_1$ & $\psi^h_2$ & $\psi^h_3$ & $\psi^h_4$ \\ \hline
 $p_x$ (\%) & 68.14  & 17.84  & 6.78  & 10.12  & 15.58  \\
 $p_y$ (\%) & 8.98  & 30.12  & 19.75  & 19.12  & 48.92    \\
 $p_z$ (\%) & 22.59  & 51.66  & 72.97  & 70.30  & 35.12    \\
 $s$ (\%)   & 0.29  & 0.38  & 0.50  & 0.46  & 0.38  \\
\hline
\end{tabular}
\end{table}

\begin{figure}[t!]
\includegraphics[width=\columnwidth]{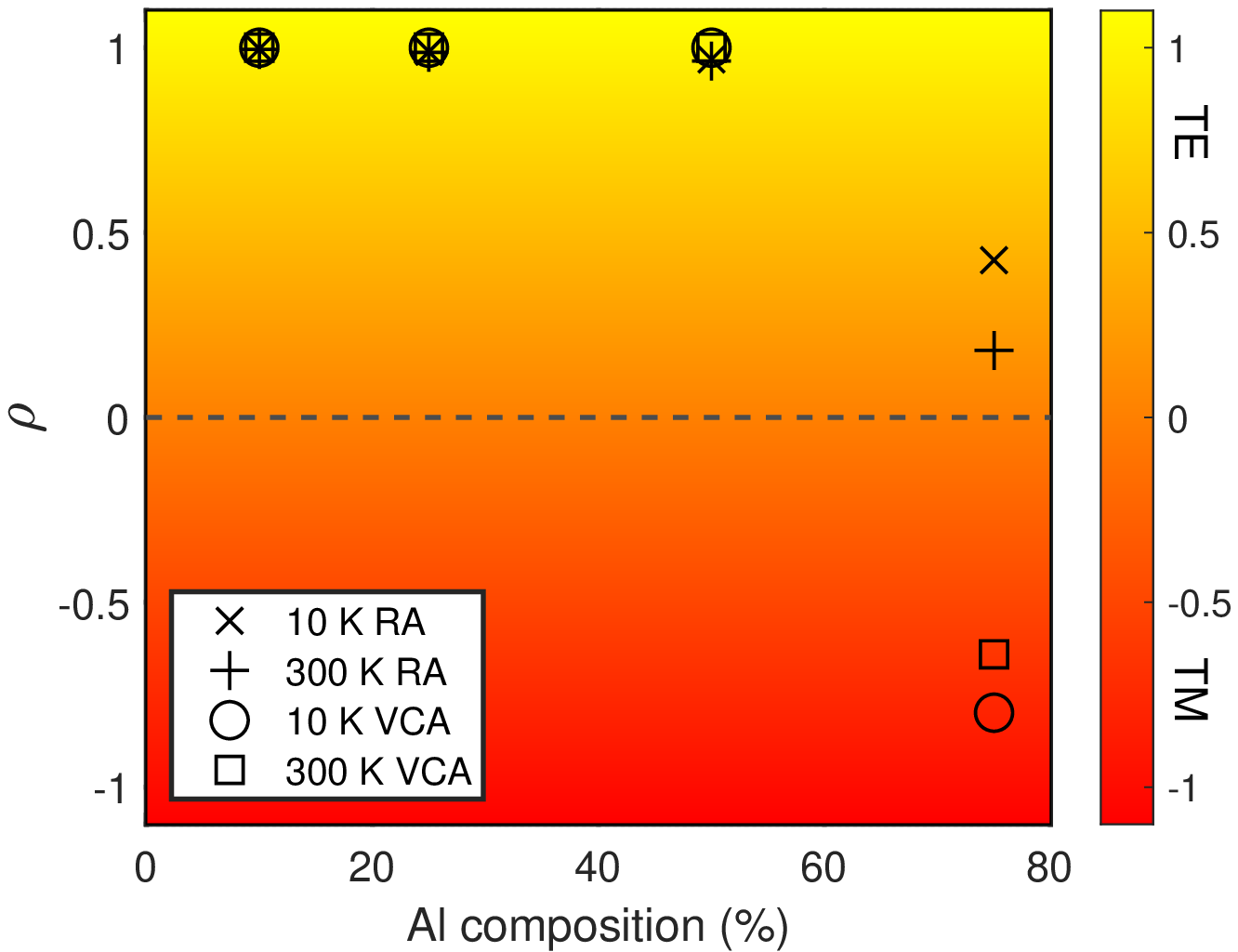}
\caption{Degree of optical polarization, $\rho$, in $c$-plane Al$_x$Ga$_{1-x}$N/AlN QWs as a function of the Al content $x$ in the well. The calculations have been performed at a fixed carrier density of \mbox{$1\times10^{18}$ cm$^{-3}$}. The degree of optical polarization has been evaluated for a low temperature ($T=10$ K) and for room temperature ($T=300$ K). Data are shown for a virtual crystal approximation (VCA) and an atomistic calculation that accounts for alloy fluctuations (RA)}.
\label{fig:DOLP}
\end{figure}

Equipped with all this information, Figure~\ref{fig:DOLP} shows the DOP, $\rho$, as a function of the Al content $x$ in the well and for two different temperatures, namely $T=10$ K and $T=300$ K; a carrier density of $1\times 10^{18}$ cm$^{-3}$ has been assumed in the calculations. Given that we are dealing here with a relatively low carrier density~\cite{FrNi2020_APL,FrNi2020_PSS}, and the fact that we are interested in a comparison between results from VCA and calculations that resolve alloy fluctuations, the impact of the screening of the built-in field is of secondary importance for our analysis, since both models make the same assumption. Future investigation can target the DOP as a function of the carrier density and thus include built-in field screening effects. 

Our calculations reveal that up to 50\% Al content, the impact of random alloy fluctuations is of secondary importance for the DOP, $\rho$, since both VCA and random alloy calculations indicate mainly TE polarized light emission. We note that for the 50\% Al case a slightly smaller $\rho$ value is found in the random alloy calculations at 300 K, but the emission will still be primarily TE polarized ($\rho\approx 1$). The situation changes noticeably for 75\% Al in the well.  
The random alloy model predicts at a temperature of $T=10$ K predominantly TE polarized light but $\rho$ is reduced ($\rho\approx 0.43$) compared to the lower Al content systems. In contrast, the VCA model predicts mainly TM polarized light emission with $\rho\approx-0.8$. The difference can be explained as follows. At low temperatures and lower carrier densities, states close to the band edge are mainly being populated. In the random alloy case these near band edge valence states are strongly localized, which still have significant contributions from $p_x$- and $p_y$-like basis states (see also discussion above). In the case of the VCA model these alloy induced localization effects are missing and band mixing effects can be different.

However, for higher temperatures ($T=300$ k), $\rho$ starts to reduce further in the random alloy case, resulting in only 'weakly' TE polarized light ($\rho\approx 0.18$). In the VCA case $\rho$ increases, even though only slightly. Nevertheless, VCA still predicts predominately TM polarized light. In general, the behavior is linked to the fact that with increasing temperature lower lying valence band states (higher lying hole states) are being populated. In the VCA case that can mean that the carriers populate valence subbands which carry a larger fraction of $p_x$ and $p_y$-like orbital character, resulting in an increase in $\rho$ or a reduction in TM polarized light with increasing temperature or carrier density. The opposite is true for the random alloy case. While the localized states near the band edge may predominantly be $p_x$ and $p_y$-like in character (e.g. $\psi^h_0$ in Table~\ref{tab:bandmix}), lower lying valence states, even if localized, may now start to contain predominantly $p_z$-like contributions, e.g. states $\psi_1$, $\psi_2$ and $\psi_3$ in Table~\ref{tab:bandmix} for the $c$-plane  Al$_{0.75}$Ga$_{0.25}$N QW. This will reduce $\rho$ and thus decrease the TE character of the emitted light.

\section{Conclusion}
\label{sec:Concl}

In this work we presented a theoretical study of the electronic and optical properties of $c$-plane (Al,Ga)N/AlN QWs.
The theoretical framework is based on an empirical atomistic TB model that accounts for (random) alloy fluctuations on a microscopic level, including connected fluctuations in strain and polarization fields. Overall, we have not only investigated the impact of random alloy fluctuations on the results but also how the electronic and optical properties change with Al content in the well. Here, we have studied $c$-plane (Al,Ga)N/AlN QWs with Al contents of 10\%, 25\%, 50\% and 75\%. This composition range is experimentally relevant for designing light emitters operating in the UV-A to UV-C spectral range.

Our analysis reveals that across the Al content investigated, random alloy fluctuations are already sufficient to lead to strong hole localization effects in $c$-plane (Al,Ga)N/AlN QWs. For electrons the situation is slightly different. At lower Al contents, e.g. 10\% Al, electron wave functions are to a first approximation delocalized across the $c$-plane. However, for higher Al contents, e.g. $>50$\%, electrons may also start to exhibit localization features. Such a situation is usually not seen in (In,Ga)N which we attribute to (i) that high In contents (e.g. $>$ 50\%) can experimentally not be reached and (ii) the effective electron masses in GaN and AlN are higher when compared to InN.       

Moreover, we find that alloy fluctuations can also affect the crossover from TE to TM polarized light with increasing Al content in the well. When comparing results from calculations resolving the alloy fluctuations with the outcome of studies performed by means of a VCA, we find that VCA predicts a crossover from TE to TM in the Al content range between 50\% and 75\%, while the random alloy calculations still predict TE polarized light. We have discussed that this effect is related to strongly localized hole states near the valence band edge which are missing in the VCA case.
Overall, while random alloy fluctuation induced carrier localization effects may have a detrimental effect on the radiative recombination rate, since electron and hole wave functions may not only be spatially separated along the growth directions but also in the growth plane as our calculations show, they may be beneficial for the degree of optical polarization. The latter aspect is extremely important for the LEE in deep UV light emitters.  

Overall, our studies show that it is key to consider random alloy fluctuations in the theoretical modelling of $c$-plane (Al,Ga)N QWs to obtain an accurate description of their electronic and optical properties. The here established model presents now an ideal starting point to study the properties of (Al,Ga)N-based optoelectronic devices, e.g. via our recently developed multi-scale transport model~\cite{MODCh2021,MODCh2022}, and ultimately to tailor their properties for future UV light emitters with improved efficiencies.

\begin{acknowledgments}
This work received funding from the Sustainable Energy Authority of Ireland and the Science Foundation Ireland (Nos. 17/CDA/4789 and 12/RC/2276 P2). 
\\ \\
The data that support the findings of this study are available from the corresponding author upon reasonable request.
\end{acknowledgments}

\bibliography{Rob_bib}

\providecommand{\noopsort}[1]{}\providecommand{\singleletter}[1]{#1}%
\begin{thebibliography}{65}%
\makeatletter
\providecommand \@ifxundefined [1]{%
 \@ifx{#1\undefined}
}%
\providecommand \@ifnum [1]{%
 \ifnum #1\expandafter \@firstoftwo
 \else \expandafter \@secondoftwo
 \fi
}%
\providecommand \@ifx [1]{%
 \ifx #1\expandafter \@firstoftwo
 \else \expandafter \@secondoftwo
 \fi
}%
\providecommand \natexlab [1]{#1}%
\providecommand \enquote  [1]{``#1''}%
\providecommand \bibnamefont  [1]{#1}%
\providecommand \bibfnamefont [1]{#1}%
\providecommand \citenamefont [1]{#1}%
\providecommand \href@noop [0]{\@secondoftwo}%
\providecommand \href [0]{\begingroup \@sanitize@url \@href}%
\providecommand \@href[1]{\@@startlink{#1}\@@href}%
\providecommand \@@href[1]{\endgroup#1\@@endlink}%
\providecommand \@sanitize@url [0]{\catcode `\\12\catcode `\$12\catcode
  `\&12\catcode `\#12\catcode `\^12\catcode `\_12\catcode `\%12\relax}%
\providecommand \@@startlink[1]{}%
\providecommand \@@endlink[0]{}%
\providecommand \url  [0]{\begingroup\@sanitize@url \@url }%
\providecommand \@url [1]{\endgroup\@href {#1}{\urlprefix }}%
\providecommand \urlprefix  [0]{URL }%
\providecommand \Eprint [0]{\href }%
\providecommand \doibase [0]{https://doi.org/}%
\providecommand \selectlanguage [0]{\@gobble}%
\providecommand \bibinfo  [0]{\@secondoftwo}%
\providecommand \bibfield  [0]{\@secondoftwo}%
\providecommand \translation [1]{[#1]}%
\providecommand \BibitemOpen [0]{}%
\providecommand \bibitemStop [0]{}%
\providecommand \bibitemNoStop [0]{.\EOS\space}%
\providecommand \EOS [0]{\spacefactor3000\relax}%
\providecommand \BibitemShut  [1]{\csname bibitem#1\endcsname}%
\let\auto@bib@innerbib\@empty
\bibitem [{\citenamefont {Brandes}\ and\ \citenamefont
  {Kettemann}(2003)}]{BrKe2003}%
  \BibitemOpen
  \bibfield  {author} {\bibinfo {author} {\bibfnamefont {T.}~\bibnamefont
  {Brandes}}\ and\ \bibinfo {author} {\bibfnamefont {S.}~\bibnamefont
  {Kettemann}},\ }\href@noop {} {\emph {\bibinfo {title} {Anderson Localization
  and Its Ramifications}}}\ (\bibinfo  {publisher} {Springer},\ \bibinfo {year}
  {2003})\BibitemShut {NoStop}%
\bibitem [{\citenamefont {Thouless}(1970)}]{Thou1970}%
  \BibitemOpen
  \bibfield  {author} {\bibinfo {author} {\bibfnamefont {D.~J.}\ \bibnamefont
  {Thouless}},\ }\bibfield  {title} {\bibinfo {title} {Anderson's theory of
  localized states},\ }\href@noop {} {\bibfield  {journal} {\bibinfo  {journal}
  {. Phys. C: Solid State Phys.}\ }\textbf {\bibinfo {volume} {3}},\ \bibinfo
  {pages} {1559} (\bibinfo {year} {1970})}\BibitemShut {NoStop}%
\bibitem [{\citenamefont {Piccardo}\ \emph {et~al.}(2017)\citenamefont
  {Piccardo}, \citenamefont {Li}, \citenamefont {Wu}, \citenamefont {Speck},
  \citenamefont {Bonef}, \citenamefont {Farrell}, \citenamefont {Filoche},
  \citenamefont {Martinelli}, \citenamefont {Peretti},\ and\ \citenamefont
  {Weisbuch}}]{PiLi2017}%
  \BibitemOpen
  \bibfield  {author} {\bibinfo {author} {\bibfnamefont {M.}~\bibnamefont
  {Piccardo}}, \bibinfo {author} {\bibfnamefont {C.-K.}\ \bibnamefont {Li}},
  \bibinfo {author} {\bibfnamefont {Y.-R.}\ \bibnamefont {Wu}}, \bibinfo
  {author} {\bibfnamefont {J.~S.}\ \bibnamefont {Speck}}, \bibinfo {author}
  {\bibfnamefont {B.}~\bibnamefont {Bonef}}, \bibinfo {author} {\bibfnamefont
  {R.~M.}\ \bibnamefont {Farrell}}, \bibinfo {author} {\bibfnamefont
  {M.}~\bibnamefont {Filoche}}, \bibinfo {author} {\bibfnamefont
  {L.}~\bibnamefont {Martinelli}}, \bibinfo {author} {\bibfnamefont
  {J.}~\bibnamefont {Peretti}},\ and\ \bibinfo {author} {\bibfnamefont
  {C.}~\bibnamefont {Weisbuch}},\ }\bibfield  {title} {\bibinfo {title}
  {{Localization landscape theory of disorder in semiconductors. II. Urbach
  tails of disordered quantum well layers}},\ }\href@noop {} {\bibfield
  {journal} {\bibinfo  {journal} {Phys. Rev. B}\ }\textbf {\bibinfo {volume}
  {95}},\ \bibinfo {pages} {144205} (\bibinfo {year} {2017})}\BibitemShut
  {NoStop}%
\bibitem [{\citenamefont {{Di Vito}}\ \emph {et~al.}(2020)\citenamefont {{Di
  Vito}}, \citenamefont {Pecchia}, \citenamefont {{Di Carlo}},\ and\
  \citenamefont {{Auf Der Maur}}}]{DiVPe2020}%
  \BibitemOpen
  \bibfield  {author} {\bibinfo {author} {\bibfnamefont {A.}~\bibnamefont {{Di
  Vito}}}, \bibinfo {author} {\bibfnamefont {A.}~\bibnamefont {Pecchia}},
  \bibinfo {author} {\bibfnamefont {A.}~\bibnamefont {{Di Carlo}}},\ and\
  \bibinfo {author} {\bibfnamefont {M.}~\bibnamefont {{Auf Der Maur}}},\
  }\bibfield  {title} {\bibinfo {title} {{Simulating random alloy effects in
  III-nitride light emitting diodes}},\ }\href@noop {} {\bibfield  {journal}
  {\bibinfo  {journal} {J. Appl. Phys.}\ }\textbf {\bibinfo {volume} {128}},\
  \bibinfo {pages} {041102} (\bibinfo {year} {2020})}\BibitemShut {NoStop}%
\bibitem [{\citenamefont {O'Donovan}\ \emph {et~al.}(2021)\citenamefont
  {O'Donovan}, \citenamefont {Luisier}, \citenamefont {O'Reilly},\ and\
  \citenamefont {Schulz}}]{ODoLu2021}%
  \BibitemOpen
  \bibfield  {author} {\bibinfo {author} {\bibfnamefont {M.}~\bibnamefont
  {O'Donovan}}, \bibinfo {author} {\bibfnamefont {M.}~\bibnamefont {Luisier}},
  \bibinfo {author} {\bibfnamefont {E.~P.}\ \bibnamefont {O'Reilly}},\ and\
  \bibinfo {author} {\bibfnamefont {S.}~\bibnamefont {Schulz}},\ }\bibfield
  {title} {\bibinfo {title} {{Impact of random alloy fluctuations on inter-well
  transport in InGaN/GaN multi-quantum well systems: an atomistic
  non-equilibrium Green's function study}},\ }\href@noop {} {\bibfield
  {journal} {\bibinfo  {journal} {J. Phys.: Condens. Matter}\ }\textbf
  {\bibinfo {volume} {33}},\ \bibinfo {pages} {045302} (\bibinfo {year}
  {2021})}\BibitemShut {NoStop}%
\bibitem [{\citenamefont {Patra}\ and\ \citenamefont
  {Schulz}(2020)}]{PaSc2020}%
  \BibitemOpen
  \bibfield  {author} {\bibinfo {author} {\bibfnamefont {S.~K.}\ \bibnamefont
  {Patra}}\ and\ \bibinfo {author} {\bibfnamefont {S.}~\bibnamefont {Schulz}},\
  }\bibfield  {title} {\bibinfo {title} {Exploring the potential of c-plane
  indium gallium nitride quantum dots for twin-photon emission},\ }\href@noop
  {} {\bibfield  {journal} {\bibinfo  {journal} {Nano Lett.}\ }\textbf
  {\bibinfo {volume} {20}},\ \bibinfo {pages} {234} (\bibinfo {year}
  {2020})}\BibitemShut {NoStop}%
\bibitem [{\citenamefont {Patra}\ and\ \citenamefont
  {Schulz}(2021)}]{PaSc2021}%
  \BibitemOpen
  \bibfield  {author} {\bibinfo {author} {\bibfnamefont {S.~K.}\ \bibnamefont
  {Patra}}\ and\ \bibinfo {author} {\bibfnamefont {S.}~\bibnamefont {Schulz}},\
  }\bibfield  {title} {\bibinfo {title} {Indium gallium nitride quantum dots:
  consequence of random alloy fluctuations for polarization entangled photon
  emission},\ }\href@noop {} {\bibfield  {journal} {\bibinfo  {journal} {Mater.
  Quantum. Technol.}\ }\textbf {\bibinfo {volume} {1}},\ \bibinfo {pages}
  {015001} (\bibinfo {year} {2021})}\BibitemShut {NoStop}%
\bibitem [{\citenamefont {Chichibu}\ \emph {et~al.}(2006)\citenamefont
  {Chichibu}, \citenamefont {Uedono}, \citenamefont {Onuma}, \citenamefont
  {Haskell}, \citenamefont {Chakraborty}, \citenamefont {Koyama}, \citenamefont
  {Fini}, \citenamefont {Keller}, \citenamefont {DenBaars}, \citenamefont
  {Speck}, \citenamefont {Mishra}, \citenamefont {Nakamura}, \citenamefont
  {Yamaguchi}, \citenamefont {Kamiyama}, \citenamefont {Amano}, \citenamefont
  {Akasaki}, \citenamefont {Han},\ and\ \citenamefont {Sota}}]{ChUe2006}%
  \BibitemOpen
  \bibfield  {author} {\bibinfo {author} {\bibfnamefont {S.~F.}\ \bibnamefont
  {Chichibu}}, \bibinfo {author} {\bibfnamefont {A.}~\bibnamefont {Uedono}},
  \bibinfo {author} {\bibfnamefont {T.}~\bibnamefont {Onuma}}, \bibinfo
  {author} {\bibfnamefont {B.~A.}\ \bibnamefont {Haskell}}, \bibinfo {author}
  {\bibfnamefont {A.}~\bibnamefont {Chakraborty}}, \bibinfo {author}
  {\bibfnamefont {T.}~\bibnamefont {Koyama}}, \bibinfo {author} {\bibfnamefont
  {P.~T.}\ \bibnamefont {Fini}}, \bibinfo {author} {\bibfnamefont
  {S.}~\bibnamefont {Keller}}, \bibinfo {author} {\bibfnamefont {S.~P.}\
  \bibnamefont {DenBaars}}, \bibinfo {author} {\bibfnamefont {J.~S.}\
  \bibnamefont {Speck}}, \bibinfo {author} {\bibfnamefont {U.~K.}\ \bibnamefont
  {Mishra}}, \bibinfo {author} {\bibfnamefont {S.}~\bibnamefont {Nakamura}},
  \bibinfo {author} {\bibfnamefont {S.}~\bibnamefont {Yamaguchi}}, \bibinfo
  {author} {\bibfnamefont {S.}~\bibnamefont {Kamiyama}}, \bibinfo {author}
  {\bibfnamefont {H.}~\bibnamefont {Amano}}, \bibinfo {author} {\bibfnamefont
  {I.}~\bibnamefont {Akasaki}}, \bibinfo {author} {\bibfnamefont
  {J.}~\bibnamefont {Han}},\ and\ \bibinfo {author} {\bibfnamefont
  {T.}~\bibnamefont {Sota}},\ }\bibfield  {title} {\bibinfo {title} {Origin of
  defect-insensitive emission probability in in-containing ({A}l,{I}n,{G}a){N}
  alloy semiconductors},\ }\href@noop {} {\bibfield  {journal} {\bibinfo
  {journal} {Nature Mater.}\ }\textbf {\bibinfo {volume} {5}},\ \bibinfo
  {pages} {810} (\bibinfo {year} {2006})}\BibitemShut {NoStop}%
\bibitem [{\citenamefont {Amano}\ \emph {et~al.}(2020)\citenamefont {Amano},
  \citenamefont {Collazo}, \citenamefont {{De Santi}}, \citenamefont
  {Einfeldt}, \citenamefont {Funato}, \citenamefont {Glaab}, \citenamefont
  {Hagedorn}, \citenamefont {Hirano}, \citenamefont {Hirayama}, \citenamefont
  {Ishii}, \citenamefont {Kashima}, \citenamefont {Kawakami}, \citenamefont
  {Kirste}, \citenamefont {Kneissl}, \citenamefont {Martin}, \citenamefont
  {Mehnke}, \citenamefont {Meneghini}, \citenamefont {Ougazzaden},
  \citenamefont {Parbrook}, \citenamefont {Rajan}, \citenamefont {Reddy},
  \citenamefont {R{\"o}mer}, \citenamefont {Ruschel}, \citenamefont {Sarkar},
  \citenamefont {Scholz}, \citenamefont {Schowalter}, \citenamefont {Shields},
  \citenamefont {Sitar}, \citenamefont {Sulmoni}, \citenamefont {Wang},
  \citenamefont {Wernicke}, \citenamefont {Weyers}, \citenamefont {Witzigmann},
  \citenamefont {Wu}, \citenamefont {Wunderer},\ and\ \citenamefont
  {Zhang}}]{AmCo2020_JPD}%
  \BibitemOpen
  \bibfield  {author} {\bibinfo {author} {\bibfnamefont {H.}~\bibnamefont
  {Amano}}, \bibinfo {author} {\bibfnamefont {R.}~\bibnamefont {Collazo}},
  \bibinfo {author} {\bibfnamefont {C.}~\bibnamefont {{De Santi}}}, \bibinfo
  {author} {\bibfnamefont {S.}~\bibnamefont {Einfeldt}}, \bibinfo {author}
  {\bibfnamefont {M.}~\bibnamefont {Funato}}, \bibinfo {author} {\bibfnamefont
  {J.}~\bibnamefont {Glaab}}, \bibinfo {author} {\bibfnamefont
  {S.}~\bibnamefont {Hagedorn}}, \bibinfo {author} {\bibfnamefont
  {A.}~\bibnamefont {Hirano}}, \bibinfo {author} {\bibfnamefont
  {H.}~\bibnamefont {Hirayama}}, \bibinfo {author} {\bibfnamefont
  {R.}~\bibnamefont {Ishii}}, \bibinfo {author} {\bibfnamefont
  {Y.}~\bibnamefont {Kashima}}, \bibinfo {author} {\bibfnamefont
  {Y.}~\bibnamefont {Kawakami}}, \bibinfo {author} {\bibfnamefont
  {R.}~\bibnamefont {Kirste}}, \bibinfo {author} {\bibfnamefont
  {M.}~\bibnamefont {Kneissl}}, \bibinfo {author} {\bibfnamefont
  {R.}~\bibnamefont {Martin}}, \bibinfo {author} {\bibfnamefont
  {F.}~\bibnamefont {Mehnke}}, \bibinfo {author} {\bibfnamefont
  {M.}~\bibnamefont {Meneghini}}, \bibinfo {author} {\bibfnamefont
  {A.}~\bibnamefont {Ougazzaden}}, \bibinfo {author} {\bibfnamefont {P.~J.}\
  \bibnamefont {Parbrook}}, \bibinfo {author} {\bibfnamefont {S.}~\bibnamefont
  {Rajan}}, \bibinfo {author} {\bibfnamefont {P.}~\bibnamefont {Reddy}},
  \bibinfo {author} {\bibfnamefont {F.}~\bibnamefont {R{\"o}mer}}, \bibinfo
  {author} {\bibfnamefont {J.}~\bibnamefont {Ruschel}}, \bibinfo {author}
  {\bibfnamefont {B.}~\bibnamefont {Sarkar}}, \bibinfo {author} {\bibfnamefont
  {F.}~\bibnamefont {Scholz}}, \bibinfo {author} {\bibfnamefont {L.~J.}\
  \bibnamefont {Schowalter}}, \bibinfo {author} {\bibfnamefont
  {P.}~\bibnamefont {Shields}}, \bibinfo {author} {\bibfnamefont
  {Z.}~\bibnamefont {Sitar}}, \bibinfo {author} {\bibfnamefont
  {L.}~\bibnamefont {Sulmoni}}, \bibinfo {author} {\bibfnamefont
  {T.}~\bibnamefont {Wang}}, \bibinfo {author} {\bibfnamefont {T.}~\bibnamefont
  {Wernicke}}, \bibinfo {author} {\bibfnamefont {M.}~\bibnamefont {Weyers}},
  \bibinfo {author} {\bibfnamefont {B.}~\bibnamefont {Witzigmann}}, \bibinfo
  {author} {\bibfnamefont {Y.-R.}\ \bibnamefont {Wu}}, \bibinfo {author}
  {\bibfnamefont {T.}~\bibnamefont {Wunderer}},\ and\ \bibinfo {author}
  {\bibfnamefont {Y.}~\bibnamefont {Zhang}},\ }\bibfield  {title} {\bibinfo
  {title} {The 2020 {UV} emitter roadmap},\ }\href@noop {} {\bibfield
  {journal} {\bibinfo  {journal} {J. Phys. D: Appl. Phys}\ }\textbf {\bibinfo
  {volume} {53}},\ \bibinfo {pages} {503001} (\bibinfo {year}
  {2020})}\BibitemShut {NoStop}%
\bibitem [{\citenamefont {Kneissl}\ \emph {et~al.}(2019)\citenamefont
  {Kneissl}, \citenamefont {Seong}, \citenamefont {Han},\ and\ \citenamefont
  {Amano}}]{KnSe2019}%
  \BibitemOpen
  \bibfield  {author} {\bibinfo {author} {\bibfnamefont {M.}~\bibnamefont
  {Kneissl}}, \bibinfo {author} {\bibfnamefont {T.-Y.}\ \bibnamefont {Seong}},
  \bibinfo {author} {\bibfnamefont {J.}~\bibnamefont {Han}},\ and\ \bibinfo
  {author} {\bibfnamefont {H.}~\bibnamefont {Amano}},\ }\bibfield  {title}
  {\bibinfo {title} {The emergence and prospects of deep-ultraviolet
  light-emitting diode technologies},\ }\href@noop {} {\bibfield  {journal}
  {\bibinfo  {journal} {Nat. Photonics}\ }\textbf {\bibinfo {volume} {13}},\
  \bibinfo {pages} {233} (\bibinfo {year} {2019})}\BibitemShut {NoStop}%
\bibitem [{\citenamefont {Frankerl}\ \emph
  {et~al.}(2020{\natexlab{a}})\citenamefont {Frankerl}, \citenamefont
  {Nippert}, \citenamefont {Hoffmann}, \citenamefont {Wang}, \citenamefont
  {Brandl}, \citenamefont {Lugauer}, \citenamefont {Zeisel}, \citenamefont
  {Hoffmann},\ and\ \citenamefont {Davies}}]{FrNi2020_JAP}%
  \BibitemOpen
  \bibfield  {author} {\bibinfo {author} {\bibfnamefont {C.}~\bibnamefont
  {Frankerl}}, \bibinfo {author} {\bibfnamefont {F.}~\bibnamefont {Nippert}},
  \bibinfo {author} {\bibfnamefont {M.~P.}\ \bibnamefont {Hoffmann}}, \bibinfo
  {author} {\bibfnamefont {H.}~\bibnamefont {Wang}}, \bibinfo {author}
  {\bibfnamefont {C.}~\bibnamefont {Brandl}}, \bibinfo {author} {\bibfnamefont
  {H.-J.}\ \bibnamefont {Lugauer}}, \bibinfo {author} {\bibfnamefont
  {R.}~\bibnamefont {Zeisel}}, \bibinfo {author} {\bibfnamefont
  {A.}~\bibnamefont {Hoffmann}},\ and\ \bibinfo {author} {\bibfnamefont
  {M.~J.}\ \bibnamefont {Davies}},\ }\bibfield  {title} {\bibinfo {title}
  {Strongly localized carriers in {Al}-rich {AlGaN/AlN} single quantum wells
  grown on sapphire substrates},\ }\href@noop {} {\bibfield  {journal}
  {\bibinfo  {journal} {J. Appl. Phys.}\ }\textbf {\bibinfo {volume} {127}},\
  \bibinfo {pages} {095701} (\bibinfo {year} {2020}{\natexlab{a}})}\BibitemShut
  {NoStop}%
\bibitem [{\citenamefont {Frankerl}\ \emph
  {et~al.}(2020{\natexlab{b}})\citenamefont {Frankerl}, \citenamefont
  {Nippert}, \citenamefont {Hoffmann}, \citenamefont {Brandl}, \citenamefont
  {Lugauer}, \citenamefont {Zeisel}, \citenamefont {Hoffmann},\ and\
  \citenamefont {Davies}}]{FrNi2020_PSS}%
  \BibitemOpen
  \bibfield  {author} {\bibinfo {author} {\bibfnamefont {C.}~\bibnamefont
  {Frankerl}}, \bibinfo {author} {\bibfnamefont {F.}~\bibnamefont {Nippert}},
  \bibinfo {author} {\bibfnamefont {M.~P.}\ \bibnamefont {Hoffmann}}, \bibinfo
  {author} {\bibfnamefont {C.}~\bibnamefont {Brandl}}, \bibinfo {author}
  {\bibfnamefont {H.-J.}\ \bibnamefont {Lugauer}}, \bibinfo {author}
  {\bibfnamefont {R.}~\bibnamefont {Zeisel}}, \bibinfo {author} {\bibfnamefont
  {A.}~\bibnamefont {Hoffmann}},\ and\ \bibinfo {author} {\bibfnamefont
  {M.~J.}\ \bibnamefont {Davies}},\ }\bibfield  {title} {\bibinfo {title}
  {Carrier dynamics in {Al}‐rich {AlGaN/AlN} quantum well structures governed
  by carrier localization},\ }\href@noop {} {\bibfield  {journal} {\bibinfo
  {journal} {Phys. Stat. Sol. (b)}\ }\textbf {\bibinfo {volume} {257}},\
  \bibinfo {pages} {2000242} (\bibinfo {year}
  {2020}{\natexlab{b}})}\BibitemShut {NoStop}%
\bibitem [{\citenamefont {Roble}\ \emph {et~al.}(2019)\citenamefont {Roble},
  \citenamefont {Patra}, \citenamefont {Massabuau}, \citenamefont {Frentrup},
  \citenamefont {Leontiadou}, \citenamefont {Dawson}, \citenamefont {Kappers},
  \citenamefont {Oliver}, \citenamefont {Graham},\ and\ \citenamefont
  {Schulz}}]{RoPa2019}%
  \BibitemOpen
  \bibfield  {author} {\bibinfo {author} {\bibfnamefont {A.~A.}\ \bibnamefont
  {Roble}}, \bibinfo {author} {\bibfnamefont {S.~K.}\ \bibnamefont {Patra}},
  \bibinfo {author} {\bibfnamefont {F.}~\bibnamefont {Massabuau}}, \bibinfo
  {author} {\bibfnamefont {M.}~\bibnamefont {Frentrup}}, \bibinfo {author}
  {\bibfnamefont {M.~A.}\ \bibnamefont {Leontiadou}}, \bibinfo {author}
  {\bibfnamefont {P.}~\bibnamefont {Dawson}}, \bibinfo {author} {\bibfnamefont
  {M.~J.}\ \bibnamefont {Kappers}}, \bibinfo {author} {\bibfnamefont {R.~A.}\
  \bibnamefont {Oliver}}, \bibinfo {author} {\bibfnamefont {D.~M.}\
  \bibnamefont {Graham}},\ and\ \bibinfo {author} {\bibfnamefont
  {S.}~\bibnamefont {Schulz}},\ }\bibfield  {title} {\bibinfo {title} {Impact
  of alloy fluctuations and coulomb effects on the electronic and optical
  properties of {$c$-plane GaN/AlGaN} quantum wells},\ }\href@noop {}
  {\bibfield  {journal} {\bibinfo  {journal} {Scientific Reports}\ }\textbf
  {\bibinfo {volume} {9}},\ \bibinfo {pages} {18862} (\bibinfo {year}
  {2019})}\BibitemShut {NoStop}%
\bibitem [{\citenamefont {Rudinsky}\ and\ \citenamefont
  {Karpov}(2020)}]{RuKa2020}%
  \BibitemOpen
  \bibfield  {author} {\bibinfo {author} {\bibfnamefont {M.~E.}\ \bibnamefont
  {Rudinsky}}\ and\ \bibinfo {author} {\bibfnamefont {S.~Y.}\ \bibnamefont
  {Karpov}},\ }\bibfield  {title} {\bibinfo {title} {Radiative and auger
  recombination constants and internal quantum efficiency of (0001) {AlGaN}
  deep-{UV} light-emitting diode structures},\ }\href@noop {} {\bibfield
  {journal} {\bibinfo  {journal} {Phys. Status Solidi A}\ }\textbf {\bibinfo
  {volume} {217}},\ \bibinfo {pages} {1900878} (\bibinfo {year}
  {2020})}\BibitemShut {NoStop}%
\bibitem [{\citenamefont {Coughlan}\ \emph {et~al.}(2015)\citenamefont
  {Coughlan}, \citenamefont {Schulz}, \citenamefont {Caro},\ and\ \citenamefont
  {OíReilly}}]{CoSc2015}%
  \BibitemOpen
  \bibfield  {author} {\bibinfo {author} {\bibfnamefont {C.}~\bibnamefont
  {Coughlan}}, \bibinfo {author} {\bibfnamefont {S.}~\bibnamefont {Schulz}},
  \bibinfo {author} {\bibfnamefont {M.~A.}\ \bibnamefont {Caro}},\ and\
  \bibinfo {author} {\bibfnamefont {E.~P.}\ \bibnamefont {OíReilly}},\
  }\bibfield  {title} {\bibinfo {title} {Band gap bowing and optical
  polarization switching in {Al$_{1-x}$Ga$_x$N} alloys},\ }\href@noop {}
  {\bibfield  {journal} {\bibinfo  {journal} {Phys. Status Solidi B}\ }\textbf
  {\bibinfo {volume} {252}},\ \bibinfo {pages} {879} (\bibinfo {year}
  {2015})}\BibitemShut {NoStop}%
\bibitem [{\citenamefont {Tanner}\ \emph {et~al.}(2018)\citenamefont {Tanner},
  \citenamefont {McMahon},\ and\ \citenamefont {Schulz}}]{TaMc2018}%
  \BibitemOpen
  \bibfield  {author} {\bibinfo {author} {\bibfnamefont {D.~S.~P.}\
  \bibnamefont {Tanner}}, \bibinfo {author} {\bibfnamefont {J.~M.}\
  \bibnamefont {McMahon}},\ and\ \bibinfo {author} {\bibfnamefont
  {S.}~\bibnamefont {Schulz}},\ }\bibfield  {title} {\bibinfo {title}
  {Interface roughness, carrier localization, and wave function overlap in
  c-plane {(In,Ga)N/GaN} quantum wells: interplay of well width, alloy
  microstructure, structural inhomogeneities, and {Coulomb effects}},\
  }\href@noop {} {\bibfield  {journal} {\bibinfo  {journal} {Phys. Rev. Appl.}\
  }\textbf {\bibinfo {volume} {10}},\ \bibinfo {pages} {034027} (\bibinfo
  {year} {2018})}\BibitemShut {NoStop}%
\bibitem [{\citenamefont {Nam}\ \emph {et~al.}(2004)\citenamefont {Nam},
  \citenamefont {Li}, \citenamefont {Nakarmi}, \citenamefont {Lin},\ and\
  \citenamefont {Jiang}}]{NaLi2004}%
  \BibitemOpen
  \bibfield  {author} {\bibinfo {author} {\bibfnamefont {K.~B.}\ \bibnamefont
  {Nam}}, \bibinfo {author} {\bibfnamefont {J.}~\bibnamefont {Li}}, \bibinfo
  {author} {\bibfnamefont {M.~L.}\ \bibnamefont {Nakarmi}}, \bibinfo {author}
  {\bibfnamefont {J.~Y.}\ \bibnamefont {Lin}},\ and\ \bibinfo {author}
  {\bibfnamefont {H.~X.}\ \bibnamefont {Jiang}},\ }\bibfield  {title} {\bibinfo
  {title} {{Unique optical properties of AlGaN alloys and related ultraviolet
  emitters}},\ }\href@noop {} {\bibfield  {journal} {\bibinfo  {journal} {Appl.
  Phys. Lett}\ }\textbf {\bibinfo {volume} {84}},\ \bibinfo {pages} {5264}
  (\bibinfo {year} {2004})}\BibitemShut {NoStop}%
\bibitem [{\citenamefont {Kolbe}\ \emph {et~al.}(2010)\citenamefont {Kolbe},
  \citenamefont {Knauer}, \citenamefont {Chua}, \citenamefont {Yang},
  \citenamefont {Einfeldt}, \citenamefont {Vogt}, \citenamefont {Johnson},
  \citenamefont {Weyers},\ and\ \citenamefont {Kneissl}}]{KoKn2010}%
  \BibitemOpen
  \bibfield  {author} {\bibinfo {author} {\bibfnamefont {T.}~\bibnamefont
  {Kolbe}}, \bibinfo {author} {\bibfnamefont {A.}~\bibnamefont {Knauer}},
  \bibinfo {author} {\bibfnamefont {C.}~\bibnamefont {Chua}}, \bibinfo {author}
  {\bibfnamefont {Z.}~\bibnamefont {Yang}}, \bibinfo {author} {\bibfnamefont
  {S.}~\bibnamefont {Einfeldt}}, \bibinfo {author} {\bibfnamefont
  {P.}~\bibnamefont {Vogt}}, \bibinfo {author} {\bibfnamefont {N.~M.}\
  \bibnamefont {Johnson}}, \bibinfo {author} {\bibfnamefont {M.}~\bibnamefont
  {Weyers}},\ and\ \bibinfo {author} {\bibfnamefont {M.}~\bibnamefont
  {Kneissl}},\ }\bibfield  {title} {\bibinfo {title} {{Optical polarization
  characteristics of ultraviolet (In)(Al)GaN multiple quantum well light
  emitting diodes}},\ }\href@noop {} {\bibfield  {journal} {\bibinfo  {journal}
  {Appl. Phys. Lett}\ }\textbf {\bibinfo {volume} {97}},\ \bibinfo {pages}
  {171105} (\bibinfo {year} {2010})}\BibitemShut {NoStop}%
\bibitem [{\citenamefont {Ryu}\ \emph {et~al.}(2013)\citenamefont {Ryu},
  \citenamefont {Choi}, \citenamefont {Choi},\ and\ \citenamefont
  {Shim}}]{RyCh2013}%
  \BibitemOpen
  \bibfield  {author} {\bibinfo {author} {\bibfnamefont {H.}~\bibnamefont
  {Ryu}}, \bibinfo {author} {\bibfnamefont {I.}~\bibnamefont {Choi}}, \bibinfo
  {author} {\bibfnamefont {H.}~\bibnamefont {Choi}},\ and\ \bibinfo {author}
  {\bibfnamefont {J.}~\bibnamefont {Shim}},\ }\bibfield  {title} {\bibinfo
  {title} {{Investigation of ligth Extraction Efficiency in AlGaN
  Deep-Ultraviolet Light-Emitting Diodes}},\ }\href@noop {} {\bibfield
  {journal} {\bibinfo  {journal} {Appl. Phys. Express}\ }\textbf {\bibinfo
  {volume} {6}},\ \bibinfo {pages} {062101} (\bibinfo {year}
  {2013})}\BibitemShut {NoStop}%
\bibitem [{\citenamefont {Schulz}\ \emph {et~al.}(2015)\citenamefont {Schulz},
  \citenamefont {Caro}, \citenamefont {Coughlan},\ and\ \citenamefont
  {O'Reilly}}]{ScCa2015}%
  \BibitemOpen
  \bibfield  {author} {\bibinfo {author} {\bibfnamefont {S.}~\bibnamefont
  {Schulz}}, \bibinfo {author} {\bibfnamefont {M.~A.}\ \bibnamefont {Caro}},
  \bibinfo {author} {\bibfnamefont {C.}~\bibnamefont {Coughlan}},\ and\
  \bibinfo {author} {\bibfnamefont {E.~P.}\ \bibnamefont {O'Reilly}},\
  }\bibfield  {title} {\bibinfo {title} {Atomistic anaylsis of the impact of
  alloy and well width fluctuations on the electronic and optical properties of
  {I}n{G}a{N}/{G}a{N} quantum wells},\ }\href@noop {} {\bibfield  {journal}
  {\bibinfo  {journal} {Phys. Rev. B}\ }\textbf {\bibinfo {volume} {91}},\
  \bibinfo {pages} {035439} (\bibinfo {year} {2015})}\BibitemShut {NoStop}%
\bibitem [{\citenamefont {Caro}\ \emph {et~al.}(2013)\citenamefont {Caro},
  \citenamefont {Schulz},\ and\ \citenamefont {O'Reilly}}]{CaSc2013local}%
  \BibitemOpen
  \bibfield  {author} {\bibinfo {author} {\bibfnamefont {M.~A.}\ \bibnamefont
  {Caro}}, \bibinfo {author} {\bibfnamefont {S.}~\bibnamefont {Schulz}},\ and\
  \bibinfo {author} {\bibfnamefont {E.~P.}\ \bibnamefont {O'Reilly}},\
  }\bibfield  {title} {\bibinfo {title} {Theory of local electric polarization
  and its relation to internal strain: impact on the polarization potential and
  electronic properties of group-{III} nitrides},\ }\href@noop {} {\bibfield
  {journal} {\bibinfo  {journal} {Phys. Rev. B}\ }\textbf {\bibinfo {volume}
  {88}},\ \bibinfo {pages} {214103} (\bibinfo {year} {2013})}\BibitemShut
  {NoStop}%
\bibitem [{\citenamefont {Schulz}\ \emph {et~al.}(2013)\citenamefont {Schulz},
  \citenamefont {Caro}, \citenamefont {Tan}, \citenamefont {Parbrook},
  \citenamefont {Martin},\ and\ \citenamefont {O'Reilly}}]{ScCa2013_Apex}%
  \BibitemOpen
  \bibfield  {author} {\bibinfo {author} {\bibfnamefont {S.}~\bibnamefont
  {Schulz}}, \bibinfo {author} {\bibfnamefont {M.~A.}\ \bibnamefont {Caro}},
  \bibinfo {author} {\bibfnamefont {L.-T.}\ \bibnamefont {Tan}}, \bibinfo
  {author} {\bibfnamefont {P.~J.}\ \bibnamefont {Parbrook}}, \bibinfo {author}
  {\bibfnamefont {R.~W.}\ \bibnamefont {Martin}},\ and\ \bibinfo {author}
  {\bibfnamefont {E.~P.}\ \bibnamefont {O'Reilly}},\ }\bibfield  {title}
  {\bibinfo {title} {Composition-dependent band gap and band-edge bowing in
  {AlInN}: A combined theoretical and experimental study},\ }\href@noop {}
  {\bibfield  {journal} {\bibinfo  {journal} {Appl. Phys. Express}\ }\textbf
  {\bibinfo {volume} {6}},\ \bibinfo {pages} {121001} (\bibinfo {year}
  {2013})}\BibitemShut {NoStop}%
\bibitem [{\citenamefont {O'Reilly}\ \emph {et~al.}(2002)\citenamefont
  {O'Reilly}, \citenamefont {Lindsay}, \citenamefont {Tomic},\ and\
  \citenamefont {Kamal-Saadi}}]{OReLi2002}%
  \BibitemOpen
  \bibfield  {author} {\bibinfo {author} {\bibfnamefont {E.~P.}\ \bibnamefont
  {O'Reilly}}, \bibinfo {author} {\bibfnamefont {A.}~\bibnamefont {Lindsay}},
  \bibinfo {author} {\bibfnamefont {S.}~\bibnamefont {Tomic}},\ and\ \bibinfo
  {author} {\bibfnamefont {M.}~\bibnamefont {Kamal-Saadi}},\ }\bibfield
  {title} {\bibinfo {title} {Tight-binding and k·p models for the electronic
  structure of {Ga(In)NAs} and related alloys},\ }\href@noop {} {\bibfield
  {journal} {\bibinfo  {journal} {Semicond. Sci. Technol.}\ }\textbf {\bibinfo
  {volume} {17}},\ \bibinfo {pages} {870} (\bibinfo {year} {2002})}\BibitemShut
  {NoStop}%
\bibitem [{\citenamefont {Li}\ and\ \citenamefont {P{\"o}tz}(1992)}]{LiPo1992}%
  \BibitemOpen
  \bibfield  {author} {\bibinfo {author} {\bibfnamefont {Z.}~\bibnamefont
  {Li}}\ and\ \bibinfo {author} {\bibfnamefont {W.}~\bibnamefont {P{\"o}tz}},\
  }\bibfield  {title} {\bibinfo {title} {Electronic density of states of
  semiconductor alloys from lattice-mismatched isovalent binary constituents},\
  }\href@noop {} {\bibfield  {journal} {\bibinfo  {journal} {Phys. Rev. B}\
  }\textbf {\bibinfo {volume} {46}},\ \bibinfo {pages} {2109} (\bibinfo {year}
  {1992})}\BibitemShut {NoStop}%
\bibitem [{\citenamefont {Boykin}\ \emph {et~al.}(2007)\citenamefont {Boykin},
  \citenamefont {Kharche}, \citenamefont {Klimeck},\ and\ \citenamefont
  {Korkusinski}}]{BoKh2007}%
  \BibitemOpen
  \bibfield  {author} {\bibinfo {author} {\bibfnamefont {T.~B.}\ \bibnamefont
  {Boykin}}, \bibinfo {author} {\bibfnamefont {N.}~\bibnamefont {Kharche}},
  \bibinfo {author} {\bibfnamefont {G.}~\bibnamefont {Klimeck}},\ and\ \bibinfo
  {author} {\bibfnamefont {M.}~\bibnamefont {Korkusinski}},\ }\bibfield
  {title} {\bibinfo {title} {Approximate bandstructures of semiconductor alloys
  from tight-binding supercell calculations},\ }\href@noop {} {\bibfield
  {journal} {\bibinfo  {journal} {J. Phys.: Condens. Matter}\ }\textbf
  {\bibinfo {volume} {19}},\ \bibinfo {pages} {036203} (\bibinfo {year}
  {2007})}\BibitemShut {NoStop}%
\bibitem [{\citenamefont {Winkelnkemper}\ \emph {et~al.}(2006)\citenamefont
  {Winkelnkemper}, \citenamefont {Schliwa},\ and\ \citenamefont
  {Bimberg}}]{WiSc2006}%
  \BibitemOpen
  \bibfield  {author} {\bibinfo {author} {\bibfnamefont {M.}~\bibnamefont
  {Winkelnkemper}}, \bibinfo {author} {\bibfnamefont {A.}~\bibnamefont
  {Schliwa}},\ and\ \bibinfo {author} {\bibfnamefont {D.}~\bibnamefont
  {Bimberg}},\ }\bibfield  {title} {\bibinfo {title} {Interrelation of
  structural and electronic properties in {In$_x$Ga$_{1-x}$N/GaN} quantum dots
  using an eigh-band k.p model},\ }\href@noop {} {\bibfield  {journal}
  {\bibinfo  {journal} {Phys. Rev. B}\ }\textbf {\bibinfo {volume} {74}},\
  \bibinfo {pages} {155322} (\bibinfo {year} {2006})}\BibitemShut {NoStop}%
\bibitem [{\citenamefont {Schulz}\ \emph {et~al.}(2010)\citenamefont {Schulz},
  \citenamefont {Badcock}, \citenamefont {Moram}, \citenamefont {Dawson},
  \citenamefont {Kappers}, \citenamefont {Humphreys},\ and\ \citenamefont
  {O’Reilly}}]{ScBa2010}%
  \BibitemOpen
  \bibfield  {author} {\bibinfo {author} {\bibfnamefont {S.}~\bibnamefont
  {Schulz}}, \bibinfo {author} {\bibfnamefont {T.~J.}\ \bibnamefont {Badcock}},
  \bibinfo {author} {\bibfnamefont {M.~A.}\ \bibnamefont {Moram}}, \bibinfo
  {author} {\bibfnamefont {P.}~\bibnamefont {Dawson}}, \bibinfo {author}
  {\bibfnamefont {M.~J.}\ \bibnamefont {Kappers}}, \bibinfo {author}
  {\bibfnamefont {C.~J.}\ \bibnamefont {Humphreys}},\ and\ \bibinfo {author}
  {\bibfnamefont {E.~P.}\ \bibnamefont {O’Reilly}},\ }\bibfield  {title}
  {\bibinfo {title} {Electronic and optical properties of nonpolar a-plane
  {GaN} quantum wells},\ }\href@noop {} {\bibfield  {journal} {\bibinfo
  {journal} {Phys. Rev. B}\ }\textbf {\bibinfo {volume} {82}},\ \bibinfo
  {pages} {125318} (\bibinfo {year} {2010})}\BibitemShut {NoStop}%
\bibitem [{\citenamefont {Yan}\ \emph {et~al.}(2009)\citenamefont {Yan},
  \citenamefont {Rinke}, \citenamefont {Scheffler},\ and\ \citenamefont
  {de~Walle}}]{YaRi2009}%
  \BibitemOpen
  \bibfield  {author} {\bibinfo {author} {\bibfnamefont {Q.}~\bibnamefont
  {Yan}}, \bibinfo {author} {\bibfnamefont {P.}~\bibnamefont {Rinke}}, \bibinfo
  {author} {\bibfnamefont {M.}~\bibnamefont {Scheffler}},\ and\ \bibinfo
  {author} {\bibfnamefont {C.~G.~V.}\ \bibnamefont {de~Walle}},\ }\bibfield
  {title} {\bibinfo {title} {Strain effects in {group-III} nitrides:
  Deformation potentials for {AlN}, {GaN}, and {InN}},\ }\href@noop {}
  {\bibfield  {journal} {\bibinfo  {journal} {Appl. Phys. Lett.}\ }\textbf
  {\bibinfo {volume} {95}},\ \bibinfo {pages} {121111} (\bibinfo {year}
  {2009})}\BibitemShut {NoStop}%
\bibitem [{\citenamefont {Bernardini}\ \emph {et~al.}(1997)\citenamefont
  {Bernardini}, \citenamefont {Fiorentini},\ and\ \citenamefont
  {Vanderbilt}}]{BeFi1997}%
  \BibitemOpen
  \bibfield  {author} {\bibinfo {author} {\bibfnamefont {F.}~\bibnamefont
  {Bernardini}}, \bibinfo {author} {\bibfnamefont {V.}~\bibnamefont
  {Fiorentini}},\ and\ \bibinfo {author} {\bibfnamefont {D.}~\bibnamefont
  {Vanderbilt}},\ }\bibfield  {title} {\bibinfo {title} {Spontaneous
  polarization and piezoelectric constants of {III-V} nitrides},\ }\href@noop
  {} {\bibfield  {journal} {\bibinfo  {journal} {Phys. Rev. B}\ }\textbf
  {\bibinfo {volume} {56}},\ \bibinfo {pages} {R10024(R)} (\bibinfo {year}
  {1997})}\BibitemShut {NoStop}%
\bibitem [{\citenamefont {Saito}\ and\ \citenamefont
  {Arakawa}(2002)}]{SaAr2002}%
  \BibitemOpen
  \bibfield  {author} {\bibinfo {author} {\bibfnamefont {T.}~\bibnamefont
  {Saito}}\ and\ \bibinfo {author} {\bibfnamefont {Y.}~\bibnamefont
  {Arakawa}},\ }\bibfield  {title} {\bibinfo {title} {Electronic structure of
  piezoelectric {In$_{0.2}$Ga$_{0.8}$N} quantum dots in {GaN} calculated using
  a tight-binding method},\ }\href@noop {} {\bibfield  {journal} {\bibinfo
  {journal} {Physica E}\ }\textbf {\bibinfo {volume} {15}},\ \bibinfo {pages}
  {169} (\bibinfo {year} {2002})}\BibitemShut {NoStop}%
\bibitem [{\citenamefont {Schuh}\ \emph {et~al.}(2012)\citenamefont {Schuh},
  \citenamefont {Barthel}, \citenamefont {Marquardt}, \citenamefont {Hickel},
  \citenamefont {Neugebauer}, \citenamefont {Czycholl},\ and\ \citenamefont
  {Jahnke}}]{ScBa2012}%
  \BibitemOpen
  \bibfield  {author} {\bibinfo {author} {\bibfnamefont {K.}~\bibnamefont
  {Schuh}}, \bibinfo {author} {\bibfnamefont {S.}~\bibnamefont {Barthel}},
  \bibinfo {author} {\bibfnamefont {O.}~\bibnamefont {Marquardt}}, \bibinfo
  {author} {\bibfnamefont {T.}~\bibnamefont {Hickel}}, \bibinfo {author}
  {\bibfnamefont {J.}~\bibnamefont {Neugebauer}}, \bibinfo {author}
  {\bibfnamefont {G.}~\bibnamefont {Czycholl}},\ and\ \bibinfo {author}
  {\bibfnamefont {F.}~\bibnamefont {Jahnke}},\ }\bibfield  {title} {\bibinfo
  {title} {Strong dipole coupling in nonpolar nitride quantum dots due to
  {Coulomb} effects},\ }\href@noop {} {\bibfield  {journal} {\bibinfo
  {journal} {Appl. Phys. Lett.}\ }\textbf {\bibinfo {volume} {100}},\ \bibinfo
  {pages} {092103} (\bibinfo {year} {2012})}\BibitemShut {NoStop}%
\bibitem [{\citenamefont {Caro}\ \emph {et~al.}(2012)\citenamefont {Caro},
  \citenamefont {Schulz},\ and\ \citenamefont {O’Reilly}}]{CaSc2012}%
  \BibitemOpen
  \bibfield  {author} {\bibinfo {author} {\bibfnamefont {M.~A.}\ \bibnamefont
  {Caro}}, \bibinfo {author} {\bibfnamefont {S.}~\bibnamefont {Schulz}},\ and\
  \bibinfo {author} {\bibfnamefont {E.~P.}\ \bibnamefont {O’Reilly}},\
  }\bibfield  {title} {\bibinfo {title} {Hybrid functional study of the elastic
  and structural properties of wurtzite and zinc-blende {group-III} nitrides},\
  }\href@noop {} {\bibfield  {journal} {\bibinfo  {journal} {Phys. Rev. B}\
  }\textbf {\bibinfo {volume} {86}},\ \bibinfo {pages} {014117} (\bibinfo
  {year} {2012})}\BibitemShut {NoStop}%
\bibitem [{\citenamefont {Plimpton}(1995)}]{Plim1995}%
  \BibitemOpen
  \bibfield  {author} {\bibinfo {author} {\bibfnamefont {S.~J.}\ \bibnamefont
  {Plimpton}},\ }\bibfield  {title} {\bibinfo {title} {Fast parallel algorithms
  for short-range molecular dynamics},\ }\href@noop {} {\bibfield  {journal}
  {\bibinfo  {journal} {Journal of Computational Physics}\ }\textbf {\bibinfo
  {volume} {117}},\ \bibinfo {pages} {1} (\bibinfo {year} {1995})}\BibitemShut
  {NoStop}%
\bibitem [{\citenamefont {Rigutti}\ \emph {et~al.}(2018)\citenamefont
  {Rigutti}, \citenamefont {Bonef}, \citenamefont {Speck}, \citenamefont
  {F.Tang},\ and\ \citenamefont {Oliver}}]{RiBo2018}%
  \BibitemOpen
  \bibfield  {author} {\bibinfo {author} {\bibfnamefont {L.}~\bibnamefont
  {Rigutti}}, \bibinfo {author} {\bibfnamefont {B.}~\bibnamefont {Bonef}},
  \bibinfo {author} {\bibfnamefont {J.}~\bibnamefont {Speck}}, \bibinfo
  {author} {\bibnamefont {F.Tang}},\ and\ \bibinfo {author} {\bibfnamefont
  {R.~A.}\ \bibnamefont {Oliver}},\ }\bibfield  {title} {\bibinfo {title} {Atom
  probe tomography of nitride semiconductors},\ }\href@noop {} {\bibfield
  {journal} {\bibinfo  {journal} {Scripta Materialia}\ }\textbf {\bibinfo
  {volume} {148}},\ \bibinfo {pages} {75} (\bibinfo {year} {2018})}\BibitemShut
  {NoStop}%
\bibitem [{\citenamefont {Pryor}\ \emph {et~al.}(1998)\citenamefont {Pryor},
  \citenamefont {Kim}, \citenamefont {Wang}, \citenamefont {Williamson},\ and\
  \citenamefont {Zunger}}]{PrKi1998}%
  \BibitemOpen
  \bibfield  {author} {\bibinfo {author} {\bibfnamefont {C.}~\bibnamefont
  {Pryor}}, \bibinfo {author} {\bibfnamefont {J.}~\bibnamefont {Kim}}, \bibinfo
  {author} {\bibfnamefont {L.}~\bibnamefont {Wang}}, \bibinfo {author}
  {\bibfnamefont {A.}~\bibnamefont {Williamson}},\ and\ \bibinfo {author}
  {\bibfnamefont {A.}~\bibnamefont {Zunger}},\ }\bibfield  {title} {\bibinfo
  {title} {Comparison of two methods for describing the strain profiles in
  quantum dots},\ }\href@noop {} {\bibfield  {journal} {\bibinfo  {journal} {J.
  Appl. Phys.}\ }\textbf {\bibinfo {volume} {83}},\ \bibinfo {pages} {2548}
  (\bibinfo {year} {1998})}\BibitemShut {NoStop}%
\bibitem [{\citenamefont {Sheerin}\ \emph {et~al.}(2021)\citenamefont
  {Sheerin}, \citenamefont {Tanner},\ and\ \citenamefont {Schulz}}]{ShTa2012}%
  \BibitemOpen
  \bibfield  {author} {\bibinfo {author} {\bibfnamefont {T.~P.}\ \bibnamefont
  {Sheerin}}, \bibinfo {author} {\bibfnamefont {D.~S.~P.}\ \bibnamefont
  {Tanner}},\ and\ \bibinfo {author} {\bibfnamefont {S.}~\bibnamefont
  {Schulz}},\ }\bibfield  {title} {\bibinfo {title} {Atomistic analysis of
  piezoelectric potential fluctuations in zinc-blende {InGaN/GaN} quantum
  wells: A stillinger-weber potential based analysis},\ }\href@noop {}
  {\bibfield  {journal} {\bibinfo  {journal} {Phys. Rev. B}\ }\textbf {\bibinfo
  {volume} {103}},\ \bibinfo {pages} {165201} (\bibinfo {year}
  {2021})}\BibitemShut {NoStop}%
\bibitem [{\citenamefont {Caro}\ \emph {et~al.}(2011)\citenamefont {Caro},
  \citenamefont {Schulz}, \citenamefont {Healy},\ and\ \citenamefont
  {O'Reilly}}]{CaSc2011}%
  \BibitemOpen
  \bibfield  {author} {\bibinfo {author} {\bibfnamefont {M.~A.}\ \bibnamefont
  {Caro}}, \bibinfo {author} {\bibfnamefont {S.}~\bibnamefont {Schulz}},
  \bibinfo {author} {\bibfnamefont {S.~B.}\ \bibnamefont {Healy}},\ and\
  \bibinfo {author} {\bibfnamefont {E.}~\bibnamefont {O'Reilly}},\ }\bibfield
  {title} {\bibinfo {title} {Built-in field control in alloyed $c$-plane
  {III-N} quantum dots and wells},\ }\href@noop {} {\bibfield  {journal}
  {\bibinfo  {journal} {J. Appl. Phys.}\ }\textbf {\bibinfo {volume} {109}},\
  \bibinfo {pages} {084110} (\bibinfo {year} {2011})}\BibitemShut {NoStop}%
\bibitem [{\citenamefont {Frankerl}\ \emph
  {et~al.}(2020{\natexlab{c}})\citenamefont {Frankerl}, \citenamefont
  {Nippert}, \citenamefont {Gomez-Iglesias}, \citenamefont {Hoffmann},
  \citenamefont {Brandl}, \citenamefont {Lugauer}, \citenamefont {Zeisel},
  \citenamefont {Hoffmann},\ and\ \citenamefont {Davies}}]{FrNi2020_APL}%
  \BibitemOpen
  \bibfield  {author} {\bibinfo {author} {\bibfnamefont {C.}~\bibnamefont
  {Frankerl}}, \bibinfo {author} {\bibfnamefont {F.}~\bibnamefont {Nippert}},
  \bibinfo {author} {\bibfnamefont {A.}~\bibnamefont {Gomez-Iglesias}},
  \bibinfo {author} {\bibfnamefont {M.~P.}\ \bibnamefont {Hoffmann}}, \bibinfo
  {author} {\bibfnamefont {C.}~\bibnamefont {Brandl}}, \bibinfo {author}
  {\bibfnamefont {H.-J.}\ \bibnamefont {Lugauer}}, \bibinfo {author}
  {\bibfnamefont {R.}~\bibnamefont {Zeisel}}, \bibinfo {author} {\bibfnamefont
  {A.}~\bibnamefont {Hoffmann}},\ and\ \bibinfo {author} {\bibfnamefont
  {M.~J.}\ \bibnamefont {Davies}},\ }\bibfield  {title} {\bibinfo {title}
  {Origin of carrier localization in {AlGaN}-based quantum well structures and
  implications for efficiency droop},\ }\href@noop {} {\bibfield  {journal}
  {\bibinfo  {journal} {Appl. Phys. Lett.}\ }\textbf {\bibinfo {volume}
  {117}},\ \bibinfo {pages} {102107} (\bibinfo {year}
  {2020}{\natexlab{c}})}\BibitemShut {NoStop}%
\bibitem [{\citenamefont {Rinke}\ \emph {et~al.}(2008)\citenamefont {Rinke},
  \citenamefont {Winkelnkemper}, \citenamefont {Qteish}, \citenamefont
  {Bimberg}, \citenamefont {Neugebauer},\ and\ \citenamefont
  {Scheffler}}]{RiWi2008}%
  \BibitemOpen
  \bibfield  {author} {\bibinfo {author} {\bibfnamefont {P.}~\bibnamefont
  {Rinke}}, \bibinfo {author} {\bibfnamefont {M.}~\bibnamefont
  {Winkelnkemper}}, \bibinfo {author} {\bibfnamefont {A.}~\bibnamefont
  {Qteish}}, \bibinfo {author} {\bibfnamefont {D.}~\bibnamefont {Bimberg}},
  \bibinfo {author} {\bibfnamefont {J.}~\bibnamefont {Neugebauer}},\ and\
  \bibinfo {author} {\bibfnamefont {M.}~\bibnamefont {Scheffler}},\ }\bibfield
  {title} {\bibinfo {title} {Consistent set of band parameters for the
  {group-III} nitrides {AlN}, {GaN}, and {InN}},\ }\href@noop {} {\bibfield
  {journal} {\bibinfo  {journal} {Phys. Rev. B}\ }\textbf {\bibinfo {volume}
  {77}},\ \bibinfo {pages} {075202} (\bibinfo {year} {2008})}\BibitemShut
  {NoStop}%
\bibitem [{\citenamefont {Pauling}(1960)}]{Paul1960}%
  \BibitemOpen
  \bibfield  {author} {\bibinfo {author} {\bibfnamefont {L.}~\bibnamefont
  {Pauling}},\ }\href@noop {} {\emph {\bibinfo {title} {The Nature of the
  Chemical Bond}}},\ \bibinfo {edition} {3rd}\ ed.\ (\bibinfo  {publisher}
  {Cornell University Press: Ithaca},\ \bibinfo {address} {New York},\ \bibinfo
  {year} {1960})\BibitemShut {NoStop}%
\bibitem [{\citenamefont {O’Reilly}\ \emph {et~al.}(2004)\citenamefont
  {O’Reilly}, \citenamefont {Lindsay},\ and\ \citenamefont
  {Fahy}}]{OReLi2004}%
  \BibitemOpen
  \bibfield  {author} {\bibinfo {author} {\bibfnamefont {E.~P.}\ \bibnamefont
  {O’Reilly}}, \bibinfo {author} {\bibfnamefont {A.}~\bibnamefont
  {Lindsay}},\ and\ \bibinfo {author} {\bibfnamefont {S.}~\bibnamefont
  {Fahy}},\ }\bibfield  {title} {\bibinfo {title} {Theory of the electronic
  structure of dilute nitride alloys: beyond the band-anti-crossing model},\
  }\href@noop {} {\bibfield  {journal} {\bibinfo  {journal} {J. Phys.: Condens.
  Matter}\ }\textbf {\bibinfo {volume} {16}},\ \bibinfo {pages} {S3257}
  (\bibinfo {year} {2004})}\BibitemShut {NoStop}%
\bibitem [{\citenamefont {Persson}\ \emph {et~al.}(2001)\citenamefont
  {Persson}, \citenamefont {da~Silva}, \citenamefont {Ahuja},\ and\
  \citenamefont {Johansson}}]{PeFe2001}%
  \BibitemOpen
  \bibfield  {author} {\bibinfo {author} {\bibfnamefont {C.}~\bibnamefont
  {Persson}}, \bibinfo {author} {\bibfnamefont {A.~F.}\ \bibnamefont
  {da~Silva}}, \bibinfo {author} {\bibfnamefont {R.}~\bibnamefont {Ahuja}},\
  and\ \bibinfo {author} {\bibfnamefont {B.}~\bibnamefont {Johansson}},\
  }\bibfield  {title} {\bibinfo {title} {{Effective electronic masses in
  wurtzite and zinc-blende GaN and AlN}},\ }\href@noop {} {\bibfield  {journal}
  {\bibinfo  {journal} {{Journal of Crystal Growth}}\ }\textbf {\bibinfo
  {volume} {231}},\ \bibinfo {pages} {397} (\bibinfo {year}
  {2001})}\BibitemShut {NoStop}%
\bibitem [{\citenamefont {Suzuki}\ and\ \citenamefont
  {Uenoyama}(1995)}]{SuUe1995}%
  \BibitemOpen
  \bibfield  {author} {\bibinfo {author} {\bibfnamefont {M.}~\bibnamefont
  {Suzuki}}\ and\ \bibinfo {author} {\bibfnamefont {T.}~\bibnamefont
  {Uenoyama}},\ }\bibfield  {title} {\bibinfo {title} {{First-principles
  calculations of effective-mass parameters of A1N and GaN}},\ }\href@noop {}
  {\bibfield  {journal} {\bibinfo  {journal} {Phys. Rev. B}\ }\textbf {\bibinfo
  {volume} {52}},\ \bibinfo {pages} {11} (\bibinfo {year} {1995})}\BibitemShut
  {NoStop}%
\bibitem [{\citenamefont {Kim}\ \emph {et~al.}(1997)\citenamefont {Kim},
  \citenamefont {Lambrecht},\ and\ \citenamefont {Segall}}]{KiLa1997}%
  \BibitemOpen
  \bibfield  {author} {\bibinfo {author} {\bibfnamefont {K.}~\bibnamefont
  {Kim}}, \bibinfo {author} {\bibfnamefont {W.}~\bibnamefont {Lambrecht}},\
  and\ \bibinfo {author} {\bibfnamefont {B.}~\bibnamefont {Segall}},\
  }\bibfield  {title} {\bibinfo {title} {{Effective masses and valence-band
  splittings in GaN and AlN}},\ }\href@noop {} {\bibfield  {journal} {\bibinfo
  {journal} {Phys. Rev. B}\ }\textbf {\bibinfo {volume} {56}},\ \bibinfo
  {pages} {12} (\bibinfo {year} {1997})}\BibitemShut {NoStop}%
\bibitem [{\citenamefont {Adelmann}\ \emph {et~al.}(2003)\citenamefont
  {Adelmann}, \citenamefont {Sarigiannidou},\ and\ \citenamefont
  {Jalabert}}]{AdSa2003}%
  \BibitemOpen
  \bibfield  {author} {\bibinfo {author} {\bibfnamefont {C.}~\bibnamefont
  {Adelmann}}, \bibinfo {author} {\bibfnamefont {E.}~\bibnamefont
  {Sarigiannidou}},\ and\ \bibinfo {author} {\bibfnamefont {D.}~\bibnamefont
  {Jalabert}},\ }\bibfield  {title} {\bibinfo {title} {Growth and optical
  properties of {GaN/AlN} quantum wells},\ }\href@noop {} {\bibfield  {journal}
  {\bibinfo  {journal} {Appl. Phys. Lett.}\ }\textbf {\bibinfo {volume} {82}},\
  \bibinfo {pages} {4154} (\bibinfo {year} {2003})}\BibitemShut {NoStop}%
\bibitem [{\citenamefont {{Al tahtamouni}}\ \emph {et~al.}(2012)\citenamefont
  {{Al tahtamouni}}, \citenamefont {Lin},\ and\ \citenamefont
  {Jiang}}]{taLi2012}%
  \BibitemOpen
  \bibfield  {author} {\bibinfo {author} {\bibfnamefont {T.~M.}\ \bibnamefont
  {{Al tahtamouni}}}, \bibinfo {author} {\bibfnamefont {J.}~\bibnamefont
  {Lin}},\ and\ \bibinfo {author} {\bibfnamefont {H.~X.}\ \bibnamefont
  {Jiang}},\ }\bibfield  {title} {\bibinfo {title} {Optical polarization in
  $c$-plane {Al}-rich {AlN/Al$_x$Ga$_{1-x}$N} single quantum wells},\
  }\href@noop {} {\bibfield  {journal} {\bibinfo  {journal} {Appl. Phys.
  Lett.}\ }\textbf {\bibinfo {volume} {101}},\ \bibinfo {pages} {042103}
  (\bibinfo {year} {2012})}\BibitemShut {NoStop}%
\bibitem [{\citenamefont {Ichikawa}\ \emph {et~al.}(2014)\citenamefont
  {Ichikawa}, \citenamefont {Iwata}, \citenamefont {Funato}, \citenamefont
  {Nagata},\ and\ \citenamefont {Kawakami}}]{IcIw2014}%
  \BibitemOpen
  \bibfield  {author} {\bibinfo {author} {\bibfnamefont {S.}~\bibnamefont
  {Ichikawa}}, \bibinfo {author} {\bibfnamefont {Y.}~\bibnamefont {Iwata}},
  \bibinfo {author} {\bibfnamefont {M.}~\bibnamefont {Funato}}, \bibinfo
  {author} {\bibfnamefont {S.}~\bibnamefont {Nagata}},\ and\ \bibinfo {author}
  {\bibfnamefont {Y.}~\bibnamefont {Kawakami}},\ }\bibfield  {title} {\bibinfo
  {title} {{High quality semipolar (1-102) AlGaN/AlN quantum wells with
  remarkably enhanced optical transition probabilities}},\ }\href@noop {}
  {\bibfield  {journal} {\bibinfo  {journal} {Appl. Phys. Lett.}\ }\textbf
  {\bibinfo {volume} {104}},\ \bibinfo {pages} {252102} (\bibinfo {year}
  {2014})}\BibitemShut {NoStop}%
\bibitem [{\citenamefont {Frankerl}(2021)}]{Fr2021}%
  \BibitemOpen
  \bibfield  {author} {\bibinfo {author} {\bibfnamefont {C.}~\bibnamefont
  {Frankerl}},\ }\emph {\bibinfo {title} {Optical Properties and Carrier
  Recombination Mechanisms in {AlGaN-based} Quantum Well Structures and
  Epitaxial Layers}},\ \href@noop {} {\bibinfo {type} {{Ph.D.} thesis}},\
  \bibinfo  {school} {Mathematik und Naturwissenschaften der Technischen
  Universit{\"a}t Berlin zur Erlangung des akademischen Grades} (\bibinfo
  {year} {2021})\BibitemShut {NoStop}%
\bibitem [{\citenamefont {Dawson}\ \emph {et~al.}(2016)\citenamefont {Dawson},
  \citenamefont {Schulz}, \citenamefont {Oliver}, \citenamefont {Kappers},\
  and\ \citenamefont {Humphreys}}]{DaSc2016}%
  \BibitemOpen
  \bibfield  {author} {\bibinfo {author} {\bibfnamefont {P.}~\bibnamefont
  {Dawson}}, \bibinfo {author} {\bibfnamefont {S.}~\bibnamefont {Schulz}},
  \bibinfo {author} {\bibfnamefont {R.~A.}\ \bibnamefont {Oliver}}, \bibinfo
  {author} {\bibfnamefont {M.~J.}\ \bibnamefont {Kappers}},\ and\ \bibinfo
  {author} {\bibfnamefont {C.~J.}\ \bibnamefont {Humphreys}},\ }\bibfield
  {title} {\bibinfo {title} {The nature of carrier localisation in polar and
  nonpolar {InGaN/GaN} quantum wells},\ }\href@noop {} {\bibfield  {journal}
  {\bibinfo  {journal} {J. Appl. Phys.}\ }\textbf {\bibinfo {volume} {119}},\
  \bibinfo {pages} {181505} (\bibinfo {year} {2016})}\BibitemShut {NoStop}%
\bibitem [{\citenamefont {Tanner}\ \emph {et~al.}(2016)\citenamefont {Tanner},
  \citenamefont {Caro}, \citenamefont {O'Reilly},\ and\ \citenamefont
  {Schulz}}]{TaCa2016_RA}%
  \BibitemOpen
  \bibfield  {author} {\bibinfo {author} {\bibfnamefont {D.~S.~P.}\
  \bibnamefont {Tanner}}, \bibinfo {author} {\bibfnamefont {M.~A.}\
  \bibnamefont {Caro}}, \bibinfo {author} {\bibfnamefont {E.~P.}\ \bibnamefont
  {O'Reilly}},\ and\ \bibinfo {author} {\bibfnamefont {S.}~\bibnamefont
  {Schulz}},\ }\bibfield  {title} {\bibinfo {title} {Random alloy fluctuations
  and structural inhomogeneities in $c$-plane {In$_x$Ga$_{1-x}$N} quantum
  wells: theory of ground and excited electron and hole states.},\ }\href@noop
  {} {\bibfield  {journal} {\bibinfo  {journal} {RSC Adv.}\ }\textbf {\bibinfo
  {volume} {6}},\ \bibinfo {pages} {64513} (\bibinfo {year}
  {2016})}\BibitemShut {NoStop}%
\bibitem [{\citenamefont {Ren}(2016)}]{Re2016}%
  \BibitemOpen
  \bibfield  {author} {\bibinfo {author} {\bibfnamefont {C.~X.}\ \bibnamefont
  {Ren}},\ }\bibfield  {title} {\bibinfo {title} {Polarisation fields in
  {III}-nitrides: effects and control},\ }\href@noop {} {\bibfield  {journal}
  {\bibinfo  {journal} {Materials Science and Technology}\ }\textbf {\bibinfo
  {volume} {32}},\ \bibinfo {pages} {418} (\bibinfo {year} {2016})}\BibitemShut
  {NoStop}%
\bibitem [{\citenamefont {Nippert}\ \emph {et~al.}(2016)\citenamefont
  {Nippert}, \citenamefont {Karpov}, \citenamefont {Callsen}, \citenamefont
  {Galler}, \citenamefont {Kure}, \citenamefont {Nenstiel}, \citenamefont
  {Wagner}, \citenamefont {Straßburg}, \citenamefont {Lugauer},\ and\
  \citenamefont {Hoffmann}}]{NiKa2016}%
  \BibitemOpen
  \bibfield  {author} {\bibinfo {author} {\bibfnamefont {F.}~\bibnamefont
  {Nippert}}, \bibinfo {author} {\bibfnamefont {S.~Y.}\ \bibnamefont {Karpov}},
  \bibinfo {author} {\bibfnamefont {G.}~\bibnamefont {Callsen}}, \bibinfo
  {author} {\bibfnamefont {B.}~\bibnamefont {Galler}}, \bibinfo {author}
  {\bibfnamefont {T.}~\bibnamefont {Kure}}, \bibinfo {author} {\bibfnamefont
  {C.}~\bibnamefont {Nenstiel}}, \bibinfo {author} {\bibfnamefont {M.~R.}\
  \bibnamefont {Wagner}}, \bibinfo {author} {\bibfnamefont {M.}~\bibnamefont
  {Straßburg}}, \bibinfo {author} {\bibfnamefont {H.-J.}\ \bibnamefont
  {Lugauer}},\ and\ \bibinfo {author} {\bibfnamefont {A.}~\bibnamefont
  {Hoffmann}},\ }\bibfield  {title} {\bibinfo {title} {Temperature-dependent
  recombination coefficients in {InGaN} light-emitting diodes: Hole
  localization, {Auger} processes, and the green gap},\ }\href@noop {}
  {\bibfield  {journal} {\bibinfo  {journal} {Appl. Phys. Lett.}\ }\textbf
  {\bibinfo {volume} {109}},\ \bibinfo {pages} {161103} (\bibinfo {year}
  {2016})}\BibitemShut {NoStop}%
\bibitem [{\citenamefont {McMahon}\ \emph {et~al.}(2020)\citenamefont
  {McMahon}, \citenamefont {Tanner}, \citenamefont {Kioupakis},\ and\
  \citenamefont {Schulz}}]{McMTa2020}%
  \BibitemOpen
  \bibfield  {author} {\bibinfo {author} {\bibfnamefont {J.~M.}\ \bibnamefont
  {McMahon}}, \bibinfo {author} {\bibfnamefont {D.~S.~P.}\ \bibnamefont
  {Tanner}}, \bibinfo {author} {\bibfnamefont {E.}~\bibnamefont {Kioupakis}},\
  and\ \bibinfo {author} {\bibfnamefont {S.}~\bibnamefont {Schulz}},\
  }\bibfield  {title} {\bibinfo {title} {Atomistic analysis of radiative
  recombination rate, {Stokes} shift, and density of states in $c$-plane
  {InGaN/GaN} quantum wells},\ }\href@noop {} {\bibfield  {journal} {\bibinfo
  {journal} {Appl. Phys. Lett.}\ }\textbf {\bibinfo {volume} {116}},\ \bibinfo
  {pages} {181104} (\bibinfo {year} {2020})}\BibitemShut {NoStop}%
\bibitem [{\citenamefont {McMahon}\ \emph {et~al.}(2022)\citenamefont
  {McMahon}, \citenamefont {Kioupakis},\ and\ \citenamefont
  {Schulz}}]{McMKi2022}%
  \BibitemOpen
  \bibfield  {author} {\bibinfo {author} {\bibfnamefont {J.~M.}\ \bibnamefont
  {McMahon}}, \bibinfo {author} {\bibfnamefont {E.}~\bibnamefont {Kioupakis}},\
  and\ \bibinfo {author} {\bibfnamefont {S.}~\bibnamefont {Schulz}},\
  }\bibfield  {title} {\bibinfo {title} {Atomistic analysis of {Auger}
  recombination in $c$-plane {(In,Ga)N/GaN} quantum wells:
  Temperature-dependent competition between radiative and nonradiative
  recombination},\ }\href@noop {} {\bibfield  {journal} {\bibinfo  {journal}
  {Phys. Rev. B}\ }\textbf {\bibinfo {volume} {105}},\ \bibinfo {pages}
  {195307} (\bibinfo {year} {2022})}\BibitemShut {NoStop}%
\bibitem [{\citenamefont {Jones}\ \emph {et~al.}(2017)\citenamefont {Jones},
  \citenamefont {Teng}, \citenamefont {Yan}, \citenamefont {Ku},\ and\
  \citenamefont {Kioupakis}}]{JoTe2017}%
  \BibitemOpen
  \bibfield  {author} {\bibinfo {author} {\bibfnamefont {C.~M.}\ \bibnamefont
  {Jones}}, \bibinfo {author} {\bibfnamefont {C.-H.}\ \bibnamefont {Teng}},
  \bibinfo {author} {\bibfnamefont {Q.}~\bibnamefont {Yan}}, \bibinfo {author}
  {\bibfnamefont {P.-C.}\ \bibnamefont {Ku}},\ and\ \bibinfo {author}
  {\bibfnamefont {E.}~\bibnamefont {Kioupakis}},\ }\bibfield  {title} {\bibinfo
  {title} {Impact of carrier localization on recombination in {InGaN} quantum
  wells and the efficiency of nitride light-emitting diodes: {Insights} from
  theory and numerical simulations},\ }\href@noop {} {\bibfield  {journal}
  {\bibinfo  {journal} {Appl. Phys. Lett.}\ }\textbf {\bibinfo {volume}
  {111}},\ \bibinfo {pages} {113501} (\bibinfo {year} {2017})}\BibitemShut
  {NoStop}%
\bibitem [{\citenamefont {Kneissl}\ \emph {et~al.}(2011)\citenamefont
  {Kneissl}, \citenamefont {Kolbe}, \citenamefont {Chua}, \citenamefont
  {Kueller}, \citenamefont {Lobo}, \citenamefont {Stellmach}, \citenamefont
  {Knauer}, \citenamefont {Rodiguez}, \citenamefont {Einfeldt}, \citenamefont
  {Yang}, \citenamefont {Johnson},\ and\ \citenamefont {Weyers}}]{KnKo2010}%
  \BibitemOpen
  \bibfield  {author} {\bibinfo {author} {\bibfnamefont {M.}~\bibnamefont
  {Kneissl}}, \bibinfo {author} {\bibfnamefont {T.}~\bibnamefont {Kolbe}},
  \bibinfo {author} {\bibfnamefont {C.}~\bibnamefont {Chua}}, \bibinfo {author}
  {\bibfnamefont {V.}~\bibnamefont {Kueller}}, \bibinfo {author} {\bibfnamefont
  {N.}~\bibnamefont {Lobo}}, \bibinfo {author} {\bibfnamefont {J.}~\bibnamefont
  {Stellmach}}, \bibinfo {author} {\bibfnamefont {A.}~\bibnamefont {Knauer}},
  \bibinfo {author} {\bibfnamefont {H.}~\bibnamefont {Rodiguez}}, \bibinfo
  {author} {\bibfnamefont {S.}~\bibnamefont {Einfeldt}}, \bibinfo {author}
  {\bibfnamefont {Z.}~\bibnamefont {Yang}}, \bibinfo {author} {\bibfnamefont
  {N.~M.}\ \bibnamefont {Johnson}},\ and\ \bibinfo {author} {\bibfnamefont
  {M.}~\bibnamefont {Weyers}},\ }\bibfield  {title} {\bibinfo {title}
  {{Advances in group III-nitride-based deep UV light-emitting diode
  technology}},\ }\href@noop {} {\bibfield  {journal} {\bibinfo  {journal}
  {Semicond. Sci. Technol.}\ }\textbf {\bibinfo {volume} {26}},\ \bibinfo
  {pages} {014036} (\bibinfo {year} {2011})}\BibitemShut {NoStop}%
\bibitem [{\citenamefont {Sakalauskas}\ \emph {et~al.}(2011)\citenamefont
  {Sakalauskas}, \citenamefont {Reuters}, \citenamefont {Khoshroo},
  \citenamefont {Kalisch}, \citenamefont {Heuken}, \citenamefont {Vescan},
  \citenamefont {R{\"o}ppischer}, \citenamefont {Cobet}, \citenamefont
  {Gobsch},\ and\ \citenamefont {Goldhahn}}]{SaRe2011}%
  \BibitemOpen
  \bibfield  {author} {\bibinfo {author} {\bibfnamefont {E.}~\bibnamefont
  {Sakalauskas}}, \bibinfo {author} {\bibfnamefont {B.}~\bibnamefont
  {Reuters}}, \bibinfo {author} {\bibfnamefont {L.~R.}\ \bibnamefont
  {Khoshroo}}, \bibinfo {author} {\bibfnamefont {H.}~\bibnamefont {Kalisch}},
  \bibinfo {author} {\bibfnamefont {M.}~\bibnamefont {Heuken}}, \bibinfo
  {author} {\bibfnamefont {A.}~\bibnamefont {Vescan}}, \bibinfo {author}
  {\bibfnamefont {M.}~\bibnamefont {R{\"o}ppischer}}, \bibinfo {author}
  {\bibfnamefont {C.}~\bibnamefont {Cobet}}, \bibinfo {author} {\bibfnamefont
  {G.}~\bibnamefont {Gobsch}},\ and\ \bibinfo {author} {\bibfnamefont
  {R.}~\bibnamefont {Goldhahn}},\ }\bibfield  {title} {\bibinfo {title}
  {Dielectric function and optical properties of quaternary {AlInGaN} alloys},\
  }\href@noop {} {\bibfield  {journal} {\bibinfo  {journal} {J. Appl. Phys.}\
  }\textbf {\bibinfo {volume} {110}},\ \bibinfo {pages} {013102} (\bibinfo
  {year} {2011})}\BibitemShut {NoStop}%
\bibitem [{\citenamefont {Vurgaftman}\ and\ \citenamefont
  {Meyer}(2003)}]{VuMe2003}%
  \BibitemOpen
  \bibfield  {author} {\bibinfo {author} {\bibfnamefont {I.}~\bibnamefont
  {Vurgaftman}}\ and\ \bibinfo {author} {\bibfnamefont {J.~R.}\ \bibnamefont
  {Meyer}},\ }\bibfield  {title} {\bibinfo {title} {Band parameters for
  nitrogen-containing semiconductors},\ }\href@noop {} {\bibfield  {journal}
  {\bibinfo  {journal} {J. Appl. Phys.}\ }\textbf {\bibinfo {volume} {94}},\
  \bibinfo {pages} {3675} (\bibinfo {year} {2003})}\BibitemShut {NoStop}%
\bibitem [{\citenamefont {Yan}\ \emph {et~al.}(2011)\citenamefont {Yan},
  \citenamefont {Rinke}, \citenamefont {Winkelnkemper}, \citenamefont {Qteish},
  \citenamefont {Bimberg}, \citenamefont {Scheffler},\ and\ \citenamefont
  {{C.~G. Van de Walle}}}]{YaRi2011}%
  \BibitemOpen
  \bibfield  {author} {\bibinfo {author} {\bibfnamefont {Q.}~\bibnamefont
  {Yan}}, \bibinfo {author} {\bibfnamefont {P.}~\bibnamefont {Rinke}}, \bibinfo
  {author} {\bibfnamefont {M.}~\bibnamefont {Winkelnkemper}}, \bibinfo {author}
  {\bibfnamefont {A.}~\bibnamefont {Qteish}}, \bibinfo {author} {\bibfnamefont
  {D.}~\bibnamefont {Bimberg}}, \bibinfo {author} {\bibfnamefont
  {M.}~\bibnamefont {Scheffler}},\ and\ \bibinfo {author} {\bibnamefont {{C.~G.
  Van de Walle}}},\ }\bibfield  {title} {\bibinfo {title} {Band parameters and
  strain effects in {ZnO} and {group-III} nitrides},\ }\href@noop {} {\bibfield
   {journal} {\bibinfo  {journal} {Semicond. Sci. Technol.}\ }\textbf {\bibinfo
  {volume} {26}},\ \bibinfo {pages} {014037} (\bibinfo {year}
  {2011})}\BibitemShut {NoStop}%
\bibitem [{\citenamefont {Schulz}\ \emph {et~al.}(2014)\citenamefont {Schulz},
  \citenamefont {Caro},\ and\ \citenamefont {O'Reilly}}]{ScCa2014}%
  \BibitemOpen
  \bibfield  {author} {\bibinfo {author} {\bibfnamefont {S.}~\bibnamefont
  {Schulz}}, \bibinfo {author} {\bibfnamefont {M.~A.}\ \bibnamefont {Caro}},\
  and\ \bibinfo {author} {\bibfnamefont {E.~P.}\ \bibnamefont {O'Reilly}},\
  }\bibfield  {title} {\bibinfo {title} {Impact of cation-based localized
  electronic states on the conduction and valence band structure of
  {Al$_{1-x}$In$_{x}$N} alloys},\ }\href@noop {} {\bibfield  {journal}
  {\bibinfo  {journal} {Appl. Phys. Lett.}\ }\textbf {\bibinfo {volume}
  {104}},\ \bibinfo {pages} {172102} (\bibinfo {year} {2014})}\BibitemShut
  {NoStop}%
\bibitem [{\citenamefont {Northrup}\ \emph {et~al.}(2012)\citenamefont
  {Northrup}, \citenamefont {Chua}, \citenamefont {Yang}, \citenamefont
  {Wunderer}, \citenamefont {Kneissl}, \citenamefont {Johnson},\ and\
  \citenamefont {Kolbe}}]{NoCh2012}%
  \BibitemOpen
  \bibfield  {author} {\bibinfo {author} {\bibfnamefont {J.~E.}\ \bibnamefont
  {Northrup}}, \bibinfo {author} {\bibfnamefont {C.~L.}\ \bibnamefont {Chua}},
  \bibinfo {author} {\bibfnamefont {Z.}~\bibnamefont {Yang}}, \bibinfo {author}
  {\bibfnamefont {T.}~\bibnamefont {Wunderer}}, \bibinfo {author}
  {\bibfnamefont {M.}~\bibnamefont {Kneissl}}, \bibinfo {author} {\bibfnamefont
  {N.~M.}\ \bibnamefont {Johnson}},\ and\ \bibinfo {author} {\bibfnamefont
  {T.}~\bibnamefont {Kolbe}},\ }\bibfield  {title} {\bibinfo {title} {Effect of
  strain and barrier composition on the polarization of light emission from
  {AlGaN/AlN} quantum wells},\ }\href@noop {} {\bibfield  {journal} {\bibinfo
  {journal} {Appl. Phys. Lett.}\ }\textbf {\bibinfo {volume} {100}},\ \bibinfo
  {pages} {021101} (\bibinfo {year} {2012})}\BibitemShut {NoStop}%
\bibitem [{\citenamefont {Ryu}(2014)}]{Ry2014}%
  \BibitemOpen
  \bibfield  {author} {\bibinfo {author} {\bibfnamefont {H.}~\bibnamefont
  {Ryu}},\ }\bibfield  {title} {\bibinfo {title} {Large enhancement of light
  extraction efficiency in {AlGaN}-based nanorod ultraviolet light-emitting
  diode structures},\ }\href@noop {} {\bibfield  {journal} {\bibinfo  {journal}
  {Nanoscale Research Letters}\ }\textbf {\bibinfo {volume} {9}},\ \bibinfo
  {pages} {58} (\bibinfo {year} {2014})}\BibitemShut {NoStop}%
\bibitem [{\citenamefont {Pernot}\ \emph {et~al.}(2010)\citenamefont {Pernot},
  \citenamefont {Kim}, \citenamefont {Fukahori}, \citenamefont {Inazu},
  \citenamefont {Fujita}, \citenamefont {Nagasawa}, \citenamefont {Hirano},
  \citenamefont {Ippommatsu}, \citenamefont {Iwaya}, \citenamefont {Kamiyama},
  \citenamefont {Akasaki},\ and\ \citenamefont {Amano}}]{PeKi2010}%
  \BibitemOpen
  \bibfield  {author} {\bibinfo {author} {\bibfnamefont {C.}~\bibnamefont
  {Pernot}}, \bibinfo {author} {\bibfnamefont {M.}~\bibnamefont {Kim}},
  \bibinfo {author} {\bibfnamefont {S.}~\bibnamefont {Fukahori}}, \bibinfo
  {author} {\bibfnamefont {T.}~\bibnamefont {Inazu}}, \bibinfo {author}
  {\bibfnamefont {T.}~\bibnamefont {Fujita}}, \bibinfo {author} {\bibfnamefont
  {Y.}~\bibnamefont {Nagasawa}}, \bibinfo {author} {\bibfnamefont
  {A.}~\bibnamefont {Hirano}}, \bibinfo {author} {\bibfnamefont
  {M.}~\bibnamefont {Ippommatsu}}, \bibinfo {author} {\bibfnamefont
  {M.}~\bibnamefont {Iwaya}}, \bibinfo {author} {\bibfnamefont
  {S.}~\bibnamefont {Kamiyama}}, \bibinfo {author} {\bibfnamefont
  {I.}~\bibnamefont {Akasaki}},\ and\ \bibinfo {author} {\bibfnamefont
  {H.}~\bibnamefont {Amano}},\ }\bibfield  {title} {\bibinfo {title} {{Improved
  Efficiency of 255-280 nm AlGaN-Based Light-Emitting Diodes}},\ }\href@noop {}
  {\bibfield  {journal} {\bibinfo  {journal} {Appl. Phys. Express}\ }\textbf
  {\bibinfo {volume} {3}},\ \bibinfo {pages} {061004} (\bibinfo {year}
  {2010})}\BibitemShut {NoStop}%
\bibitem [{\citenamefont {O’Donovan}\ \emph {et~al.}(2021)\citenamefont
  {O’Donovan}, \citenamefont {Chaudhuri}, \citenamefont {Streckenbach},
  \citenamefont {Farrell}, \citenamefont {Schulz},\ and\ \citenamefont
  {Koprucki}}]{MODCh2021}%
  \BibitemOpen
  \bibfield  {author} {\bibinfo {author} {\bibfnamefont {M.}~\bibnamefont
  {O’Donovan}}, \bibinfo {author} {\bibfnamefont {D.}~\bibnamefont
  {Chaudhuri}}, \bibinfo {author} {\bibfnamefont {T.}~\bibnamefont
  {Streckenbach}}, \bibinfo {author} {\bibfnamefont {P.}~\bibnamefont
  {Farrell}}, \bibinfo {author} {\bibfnamefont {S.}~\bibnamefont {Schulz}},\
  and\ \bibinfo {author} {\bibfnamefont {T.}~\bibnamefont {Koprucki}},\
  }\bibfield  {title} {\bibinfo {title} {From atomistic tight-binding theory to
  macroscale drift–diffusion: {Multiscale} modeling and numerical simulation
  of uni-polar charge transport in {(In,Ga)N} devices with random
  fluctuations},\ }\href@noop {} {\bibfield  {journal} {\bibinfo  {journal} {J.
  Appl. Phys.}\ }\textbf {\bibinfo {volume} {130}},\ \bibinfo {pages} {065702}
  (\bibinfo {year} {2021})}\BibitemShut {NoStop}%
\bibitem [{\citenamefont {O’Donovan}\ \emph {et~al.}(2022)\citenamefont
  {O’Donovan}, \citenamefont {Farrell}, \citenamefont {Streckenbach},
  \citenamefont {Koprucki},\ and\ \citenamefont {Schulz}}]{MODCh2022}%
  \BibitemOpen
  \bibfield  {author} {\bibinfo {author} {\bibfnamefont {M.}~\bibnamefont
  {O’Donovan}}, \bibinfo {author} {\bibfnamefont {P.}~\bibnamefont
  {Farrell}}, \bibinfo {author} {\bibfnamefont {T.}~\bibnamefont
  {Streckenbach}}, \bibinfo {author} {\bibfnamefont {T.}~\bibnamefont
  {Koprucki}},\ and\ \bibinfo {author} {\bibfnamefont {S.}~\bibnamefont
  {Schulz}},\ }\bibfield  {title} {\bibinfo {title} {Multiscale simulations of
  uni-polar hole transport in {(In, Ga)N} quantum well systems},\ }\href@noop
  {} {\bibfield  {journal} {\bibinfo  {journal} {Opt. Quant. Electron.}\
  }\textbf {\bibinfo {volume} {54}},\ \bibinfo {pages} {405} (\bibinfo {year}
  {2022})}\BibitemShut {NoStop}%
\end{thebibliography}%

\end{document}